\documentclass[manuscript,screen,nonacm]{acmart}

\makeatletter
\@twosidefalse      % disable two-sided mode
\@mparswitchfalse   % disable margin switching
\renewcommand\@titlefont{\fontsize{14}{16}\selectfont\bfseries} % bigger title
\makeatother

\AtBeginDocument{%
  }

\renewcommand\footnotetextcopyrightpermission[1]{} 
\usepackage{xcolor}
\usepackage{eso-pic}

\begin{document}

\title[Talking to an AI Mirror: Designing Self-Clone Chatbots for Enhanced Engagement in Mental Health Support]{Talking to an AI Mirror: Designing Self-Clone Chatbots for Enhanced Engagement in Digital Mental Health Support}

\author{Mehrnoosh Sadat Shirvani}
\affiliation{%
  \institution{University of British Columbia}
  \city{Vancouver}
  \state{British Columbia}
  \country{Canada}}
\email{mehrshi@cs.ubc.ca}

\author{Jackie Liu}
\affiliation{%
  \institution{University of British Columbia}
  \city{Vancouver}
  \state{British Columbia}
  \country{Canada}}
\email{anjieliu@cs.ubc.ca}

\author{Thomas Chao}
\affiliation{%
\institution{University of British Columbia}
  \city{Vancouver}
  \state{British Columbia}
  \country{Canada}}
\email{thomas.chao@ubc.ca}

\author{Suky Martinez}
\affiliation{%
\institution{Johns Hopkins University School of Medicine}
  \city{Baltimore}
  \state{Maryland}
  \country{USA}}
\email{smart209@jh.edu}

\author{Laura Brandt}
\affiliation{%
\institution{City College of New York}
  \city{New York}
  \state{New York}
  \country{USA}}
\email{lbrandt@ccny.cuny.edu}

\author{Ig-Jae Kim}
\affiliation{%
  \institution{Korea Institute of Science and Technology}
  \city{Seoul}
  \country{Korea}
}
\email{drjay@kist.re.kr}

\author{Dongwook Yoon}
\affiliation{%
  \institution{University of British Columbia}
  \city{Vancouver}
  \state{British Columbia}
  \country{Canada}}
\email{yoon@cs.ubc.ca}

\renewcommand{\shortauthors}{Shirvani et al.}

\begin{abstract}
\emph{\textbf{ABSTRACT}} Mental health conversational agents have the potential to deliver valuable therapeutic impact, but low user engagement remains a critical barrier hindering their efficacy. Existing therapeutic approaches have leveraged clients' internal dialogues (e.g., journaling, talking to an empty chair) to enhance engagement through accountable, self-sourced support. Inspired by these, we designed novel AI-driven self-clone chatbots that replicate users' support strategies and conversational patterns to improve therapeutic engagement through externalized meaningful self-conversation. Validated through a semi-controlled experiment (N=180), significantly higher emotional and cognitive engagement was demonstrated with self-clone chatbots than a chatbot with a generic counselor persona. Our findings highlight self-clone believability as a mediator and emphasize the balance required in maintaining convincing self-representation while creating positive interactions. This study contributes to AI-based mental health interventions by introducing and evaluating self-clones as a promising approach to increasing user engagement, while exploring implications for their application in mental health care.

\end{abstract}


\maketitle

\fancyfoot{} 
\renewcommand\footnotetextcopyrightpermission[1]{}

\section{Introduction}
Mental health issues are a pressing global concern, affecting millions of lives and significantly impacting overall well-being and quality of life \cite{rehmGlobalBurdenDisease2019, HttpsIriswhointBitstream}. Yet access to care remains limited by workforce shortages and fragmented services \cite{kakumaHumanResourcesMental2011, AccessCareData}. In response, digital solutions, particularly chatbot-based interventions, have attracted growing interest as low-barrier, scalable supports \cite{abd-alrazaqOverviewFeaturesChatbots2019}. Early studies show that well-designed conversational agents can induce deep self-disclosure and allow socially-mediated help seeking through the use of Artificial Intelligence (AI) \cite{haque2022app,lee2020designing}. Recent works in the domain of human-computer interaction (HCI) demonstrates that specific design choices, such as conceptual metaphors, persona construction, and playful interactions, shape users' perceptions of the warmth, competence, and moral agency of a chatbot, thus narrowing the experiential gap between human–human and human–AI encounters \cite{khadpe2020conceptual,wu2019agent,nikghalb2024interrogating,wester2024chatbot}. Researchers have embedded principles from established therapeutic methods, such as cognitive behavioral therapy (CBT) and mindfulness, into conversational agents driven by AI to offer mental and emotional support, such as MindfulDiary \cite{kimMindfulDiaryHarnessingLarge2024} and Sunny \cite{narainPromotingWellbeingSunny2020}. The role and potential of these digital tools for mental health care are becoming increasingly recognized by national initiatives like the WHO \cite{DigitalMentalHealth, WHO2021}.

However, the challenges in maintaining a consistent user \emph{engagement} - the degree to which a person invests effort, attention, and motivation in activities - has been extensively studied and is recognized as a key factor limiting the efficacy of digital mental health interventions.\cite{ganEffectEngagementDigital2021,torousClinicalReviewUser2018, wilhelmsenMotivationPersistInternetbased2013}. While greater engagement does not guarantee better mental health outcomes, it can promote user adherence, helping ensure that interventions are used long enough for their therapeutic benefits to be fully realized \cite{donkin2011systematic, lipschitz2023engagement}. Without sufficient client engagement, intervention outcomes can be limited, leading to minimal or inconsistent outcomes\cite{gan2021effect}. As mental health concerns persist, exacerbated by events like the COVID-19 pandemic \cite{lattieOverviewRecommendationsMore2022}, there is a need to refine these often one-size-fits-all digital tools to address the complex nature of user engagement for their therapeutic efficacy.

As digital tools evolve, \emph{self-dialogue} offers a natural foundation for designing AI systems that engage users in constructive inner conversations. Traditional techniques such as reflective journaling or the empty chair exercise of Gestalt therapy have used conversing with imagined figures or internal parts of oneself as a tool for self-reflection and personal growth \cite{perls1969gestalt, WritingEmotionalExperiences}. Across disciplines, motivational and positive self-talk has been shown to be a trainable self-regulation strategy that improves self-efficacy and confidence \cite{ryan2024self, hatzigeorgiadis2009mechanisms, moser2017third, rogelberg2013executive}, which in turn can predict motivation and adherence to health and mental health plans \cite{peters2018designing, martin2017motivation}. Consistent with Self-Determination Theory, autonomous motivation and perceived competence mediate behavior change and strengthen engagement \cite{jochems2017motivation, sheeran2020self,piltch2016role, jones2013peer}. Furthermore, building on Rogers’ view that individuals have an inherent capacity for self-healing \cite{hoffmanHumanisticPsychologySelfAcceptance2013}, one's inner dialogue can be used to nurture this capacity for emotional healing \cite{hardySpeakingClearlyCritical2006, krossSelftalkRegulatoryMechanism2014, todEffectsSelfTalkSystematic2011}. While promising, integrating self-talk into digital tools requires careful design, as context and users' mindsets can determine whether dialogue supports constructive change through creating emotional distancing or may risk amplifying negative emotions \cite{krossSelftalkRegulatoryMechanism2014}.

Whereas most digital health tools lean toward merely digitizing established methods, generative AI can now enable genuinely new approaches to personalized care. An emerging trend in HCI is the concept of \emph{clones}, which utilize personal data to create digital replicas of individuals \cite{mcilroy-youngMimeticModelsEthical2022,slaterExperimentalStudyVirtual2019,leeSpeculatingRisksAI2023}, and aligns well with re-framing self-management as collaboration with oneself through AI \cite{burgess2022just}. Early explorations of digital twins in forms of communicating with the future self \cite{pataranutaporn2024future, jeon2025letters}, large-language-model-based human simulations~\cite{wang2025limits}, and generative agent simulations \cite{park2024generative} suggest that such framings can prompt more thoughtful engagement with AI explanations \cite{buccinca2021trust} and hold significant implications for introspection and self-sourced emotional support. A striking illustration is Michelle Huang's experience using GPT-3 and the content of her childhood journals to create a conversational "time portal" with her younger self, reporting experiencing unexpected emotional healing \cite{michellehuang[@michellehuang42]TrainedAiChatbot2022}. 

Motivated by these, we aimed to design \emph{self-clone} chatbots, which enable the externalization of inner dialogues, and explore their potential to enhance user engagement with AI-based mental health chatbots, thereby fostering an environment necessary for the optimal delivery of mental health support. Our study is guided by the following research questions (RQs):

\begin{itemize}
\item RQ1: How can we design a self-clone chatbot to engage users effectively?
\item RQ2: To what extent can the self-clone persona help engagement levels?
\item RQ3: How do users perceive self-clones, and what factors shape the experience?
\end{itemize}

Drawing from existing human-AI design guidelines and evidence on common engagement barriers, we established three design principles for the self-clone chatbots to guide our process: 1) counseling goal and approach, 2) persona, and 3) selective trait representation. We brainstormed with an interdisciplinary team, including clinical psychologists, and refined ten design scenarios for potential theoretical framing of the self-clones, ultimately selecting the most promising one for implementation and evaluation. Based on our design principles, our chatbot is informed by the empty chair technique\cite{hoffmanHumanisticPsychologySelfAcceptance2013}, utilizing a future-self persona, and focuses on recreating conversational and support patterns. For data collection, we developed a novel approach using a hypothetical chatbot that simulates how users would support a friend in need, allowing us to capture users' natural social support strategies and behavioral patterns. The captured support behavior was incorporated into our self-clones through our new prompting technique, social support prompting (SSP). Informed by psychological theories on types of supporting strategies, SSP shows promise to improve the model's ability to accurately identify and replicate specific social support patterns observed in the data.

To evaluate the impact of our self-clone chatbot designs on user engagement, we conducted a semi-controlled experiment with 180 participants, randomly assigned across three conditions: baseline (BL), self-clone without SSP (SCX), and self-clone with SSP (SCS). SCX was designed to replicate a user’s behavioral and conversational style, such as tone and word choice, while SCS, utilizing SSP, also incorporated their preferred supportive strategies. Statistical analysis of the data showed a significant enhancement in cognitive and emotional engagement compared to the baseline. Notably, SCS demonstrated significantly higher emotional engagement toward the baseline than SCX. Moreover, self-clone believability emerged as an important mediating factor. Additionally, thematic analysis of participants' comments revealed that this believability was likely influenced by participants' perceptions toward themselves and the inherent limitations of the 'talking across timelines' concept (i.e., talking to the future self). Further statistical and thematic analyses indicated generally positive perceptions, including higher motivation and acceptance among participants.

Our work contributes: 1) the design of self-clone chatbots as a novel AI-driven approach, along with SSP as a promising technique to enhance the capabilities of self-clones in recreating user support strategies, 2) evidence of improved user engagement, and 3) insights into users' perceptions of the self-clone chatbots alongside the implications for future applications.

\section{Related Works} 
\label{sec:rw}
\subsection{The Role of Self in Mental Health}
In psychology, the concept of \emph{self} is not merely an abstract notion but a fundamental element in therapeutic interventions. Each individual constructs a personal narrative based on their beliefs, experiences, and perceptions, which collectively shape their identity \cite{polkinghorneNarrativeSelfConcept1991}. This self-narrative plays a critical role in mental well-being, serving as the lens through which individuals interpret their experiences and interact with the world \cite{hardySpeakingClearlyCritical2006, olesTypesInnerDialogues2020}. A key aspect of this narrative is self-talk or inner dialogue --- a form of self-directed verbal expression unique to each individual \cite{polkinghorneNarrativeSelfConcept1991, joNeuralEffectsOnes2024, brinthauptIndividualDifferencesSelfTalk2019}. While positive self-talk can foster motivation, resilience, and emotional regulation, dysfunctional self-talk is often present in psychological disorders such as depression, anxiety, and personality disorders \cite{todEffectsSelfTalkSystematic2011}.

Addressing self-talk in a therapeutic setting can help reconstruct personal narratives more positively and foster resilience. Many therapeutic approaches aim to improve self-perception by enhancing the quality of self-talk. For example, a key component of Carl Rogers' person-centered therapy is to shift inner dialogues toward a more accepting form, aligning the real self (who one actually is) with the ideal self (who one wants to be) \cite{rogersClientcenteredTherapyIts1951, velasquezReapproachingRogersDiscursive2018}. Other therapeutic techniques, such as Internal Family Systems (IFS), Cognitive Behavioral Therapy (CBT), and Compassion-Focused Therapy (CFT), also leverage self-talk to increase self-compassion, address distorted self-perceptions, and resolve internal conflicts \cite{earleySelfTherapyStepStepGuide2009, grzybowskiTraitMindfulnessSelfCompassion2022, brinthauptSelftalkResearchChallenges2023}. Research shows that creating psychological distance from one's inner thoughts---self-distancing---can enhance self-awareness and improve emotional regulation \cite{gainsburgDistancedSelftalkChanges2020, krossChapterTwoSelfDistancing2017}. While externalizing self-talk through techniques like journaling and expressive writing can promote distancing from negative internal “loops”, these methods are often self-guided and non-interactive, highlighting an opportunity for more dynamic approaches to facilitate this process \cite{parkSteppingBackMove2016, ruiniWritingTechniquePsychotherapies2022}.

\subsection{AI Clones and Digital Twins}
With the recent advancement in AI, there is an emerging trend of human digital replicas, often called digital twins or clones. From deepfakes to voice-clone, the recreation of famous personalities through mimicking visual or audio likenesses has become popular in online communities \cite{westerlundEmergenceDeepfakeTechnology2019, ruggieroVoiceCloningMultiSpeaker2021a}. Using generative AI, platforms such as character.ai have emerged to offer beyond access to pre-programmed chatbots of famous individuals, enabling users to create their own clones of different people \cite{CharacteraiPersonalizedAI}. Going beyond entertainment, there has also been a rise in task delegation clones, with platforms such as personal.ai or fraia.ai offering to create clones of users to perform the role of real-world counterparts without real-world costs \cite{PersonalAI, FraiaAI}.

Although the term \emph{digital twin} has recently been adapted to refer to digital clones in online settings, the concept originated in industrial engineering as a digital counterpart designed to mimic real-world entities in a virtual environment, primarily for optimizing complex control systems \cite{v.vPersonalDigitalTwins2020}. Today, however, the use of digital twins has expanded across many domains. In the health sector, digital twins of patients, created from health and biology data and symptoms, are used for preventive measures, such as illness identification, personalized diagnoses, and treatments \cite{sunDigitalTwinHealthcare2023}. In marketing, Truby and Brown \cite{trubyHumanDigitalThought2021} developed what they called a \emph{human digital thought clone}, using customer data to replicate consumer thought processes and decision-making for profit maximization. Within HCI, McIlroy-Young et al. \cite{mcilroy-youngMimeticModelsEthical2022} introduced the concept of \emph{mimetic models}, which use personal data to train AI systems that mimic an individual. Similarly, Lee et al. \cite{leeSpeculatingRisksAI2023} define AI clones as interactive digital representations of individuals, trained on personal data, that replicate certain traits or behaviors of the person. Our study expands upon the existing research in the field of AI clones by examining potential use cases for these clones when utilized by the original individual. We refer to these clones as self-clones.

Although AI clones have not been explicitly studied for emotional support, there are relevant cases that suggest their potential. For instance, Replika offers an AI companion that can be personalized to users' preferences, with some individuals using this feature to create clones of real-world people they know, substituting for lack of real-world relationships, and many have reported finding comfort and emotional support through these interactions \cite{Lennig2023, Cox2023}. Slater et al. \cite{slaterExperimentalStudyVirtual2019} explored the therapeutic implications of conversing with a static self-avatar in virtual reality environments, finding that simulated self-conversations led to greater perceived helpfulness. Similarly, our study focuses on the implications of using a digital representation of oneself for mental health benefits. However, we specifically explore using AI to enable interactivity, simulating natural, human-like conversations through a chatbot.

\subsection{AI-Driven Mental Health Interventions}
Digital mental health interventions (DMHIs) refer to the use of technology-based tools and digital platforms to deliver mental health support, which can take various forms, such as mobile applications, web-based programs, or video games \cite{parkFrameworkImplementationDigital}. The growing demand for accessible mental health care, coupled with recent advancements in AI capabilities, has the potential to transform the development of DMHIs. For example, AI-driven chatbots have emerged as a promising tool for screening and symptom management \cite{boucherArtificiallyIntelligentChatbots2021}. Recent studies have explored the potential of receiving mental health benefits by supporting an AI chatbot \cite{kim2020helping}, and conversational agents such as Woebot \cite{fitzpatrickDeliveringCognitiveBehavior2017}, Wysa \cite{inksterEmpathyDrivenConversationalArtificial2018}, and Tess \cite{fulmerUsingPsychologicalArtificial2018} have gained popularity for managing mild to moderate symptoms, demonstrating the therapeutic potential of these tools \cite{casuAIChatbotsMental2024}.

DMHIs are closely integrated with digital systems, making HCI principles valuable for designing safe and effective mental health tools \cite{balcombe2022human}. Recent research on designing AI-driven chatbots emphasizes the importance of a consistent chatbot personality, transparency, designing for dialog failures, and fostering human-like conversations \cite{jainEvaluatingInformingDesign2018}. Additionally, guidelines for human-AI interactions (HAI) highlight the need to account for potential biases and align with relevant social norms based on the user’s context \cite{amershiGuidelinesHumanAIInteraction2019}. Our study incorporates these design practices and guidelines to create a thoughtful, user-centered approach to developing AI-driven mental health tools.

\subsection{User Engagement in Digital Mental Health Tools} 
User engagement (also referred to as client engagement) is a critical and frequently studied factor in digital mental health interventions, as higher levels of engagement are strongly associated with greater therapeutic effectiveness \cite{ganEffectEngagementDigital2021, yardleyUnderstandingPromotingEffective2016}. Although engagement is not directly linked to improved mental health, it is essential for sustaining involvement long enough for the intervention to be fully delivered. Engagement is a complex, multi-dimensional process, and while there is not yet a universally accepted definition, it is commonly categorized into three main components: behavioral, cognitive, and emotional engagement. Improvements in any of these areas are thought to enhance overall engagement \cite{bijkerkMeasuringEngagementMental2023}. Behavioral engagement is typically characterized by intervention adherence and clients' active efforts while using the tool \cite{perskiConceptualisingEngagementDigital2017}. In HCI research, it is often measured through objective metrics such as the number of clicks, time spent per task, or frequency of use \cite{obrienWhatUserEngagement2008}. Cognitive engagement refers to the degree to which clients agree with the therapeutic approach, meaning how suitable they perceive the intervention to be in helping them reach their goals \cite{yardleyUnderstandingPromotingEffective2016}. Emotional engagement, or affective engagement, reflects the user's subjective experience with the tool, including interest, emotional connection, and overall satisfaction \cite{perskiConceptualisingEngagementDigital2017}. A review of recent studies reveals that most research in the HCI and DMHI domains employs a combination of adapted self-report or observer-rated questionnaires alongside objective measures, which aligns with the recommendation to use mixed methods to comprehensively assess the complexity of engagement in relation to specific study contexts \cite{perskiSelfreportMeasureEngagement2020}.

User engagement can be influenced by a variety of factors. Common contributors to low engagement include a lack of motivation to persist with self-guided interventions, poor usability and user experience, and a misalignment between the provided support and the user's needs \cite{torousClinicalReviewUser2018, wilhelmsenMotivationPersistInternetbased2013}. Other barriers include user-related factors (e.g., demographic variables, beliefs, prior experience, and skills), program-related factors (e.g., perceived usefulness and fit), and technology- and environment-related factors (e.g., privacy concerns and technical issues) \cite{borghoutsBarriersFacilitatorsUser2021}.  Studies have shown that HCI practices, including user-centric design, persuasive design practices, gamification, and external motivators (e.g., reminders), can improve engagement \cite{wildeboerRelationshipPersuasiveTechnology2016, brownGamificationAdherenceWebBased2016}. However, these approaches often fail to address the root issue—the one-size-fits-all therapeutic approach commonly used in digital interventions \cite{AdaptiveSystemsInternetDelivered, hornsteinPersonalizationStrategiesDigital2023}. This challenge, along with the aforementioned barriers, highlights the need for a more adaptive and personalized approach to mental health support, which would leverage individual differences to be better tailored to meet the specific traits and needs of users.

\section{Designing Self-Clone Chatbots}
One of our primary objectives is to design a self-clone chatbot for mental health support that maximizes user engagement. Grounded in existing literature and best practices for conversational agent design, we focused on three key principles related to our topic. Using these insights, we brainstormed and iteratively refined high-impact use case scenarios (see Appendix \ref{app:desscen}), incorporating feedback from our advisory board of three clinical psychologists during weekly meetings. Ultimately, we selected one scenario that, in our team's opinion, best illustrated the potential of self-clone technology in mental health support. This scenario was implemented using two distinct approaches: one focused on replicating the user’s conversational style and the other on enhancing it with an added layer of replicating the supportive strategies through a novel technique we refer to as Social Support Prompting.

\subsection{Design Principles}
\label{sec:despri}
Designing self-clones to enhance engagement for mental health support necessitates a careful approach to addressing interdisciplinary needs. Our overarching strategy is informed by Peter et al.’s Responsible Design Process \cite{peters2020responsible}, and we draw upon insights from the existing literature on human-AI interactions, the use of AI conversational agents in digital mental health care, and the obstacles to engagement in digital mental health interventions mentioned in section \ref{sec:rw}. In collaboration with our interdisciplinary team, we adapted established guidelines, including Amershi et al.'s HAI guidelines \cite{amershiGuidelinesHumanAIInteraction2019} and Ahmad et al.'s guidelines for conversational agents in mental health \cite{ahmad2022designing}. These were assessed, consolidated, and modified to meet the specific needs of self-clones, resulting in three key design principles tailored to the specific design challenges of self-clone. The following section details these principles and their role in our chatbot's design.

\begin{itemize}
\item {\textbf{Leveraging Promising Counseling Goal and Approach}}: 

Our first design principle combines two key elements of clear goal-setting with proper theoretical grounding. Goal setting is important in therapy to provide direction and measurable outcomes \cite{epton2017unique, tryon2018meta}, and in human-AI design to set expectations and enhance user autonomy \cite{amershiGuidelinesHumanAIInteraction2019}. As a lack of theoretical grounding in counseling and therapeutic interventions and a clear goal can limit the effectiveness of mental health chatbots \cite{boucherArtificiallyIntelligentChatbots2021}, this calls for an informed approach rooted in established counseling methodologies. Such grounding ensures the chatbot’s perceived competence \cite{ahmad2022designing} and supports the effective delivery of overarching therapeutic techniques. While the specific goal and AI’s behavioral rules may vary depending on the chosen therapeutic framework, to create a supportive environment, the self-clone chatbot should adhere to established principles for fostering a digital therapeutic alliance. At its core, the system should demonstrate empathy, reflect similar emotions, and validate the user’s feelings \cite{thieme2023designing, morris2018towards}. Leveraging AI’s strengths, the self-clone can serve as an interactive alternative to self-guided methods like reflective journaling or dynamically adapt care delivery to users’ cognitive and emotional needs, enabling tailored conversational styles and engaging therapeutic experiences \cite{ahmad2022designing, kocaballi2020conversational}.

\item {\textbf{Having a Familiar yet Distinct Persona}}: Recent studies demonstrate a well-crafted persona gives a chatbot a consistent voice and behavior, which helps build trust and create a more engaging user experience \cite{chavesHowShouldMy2021,ruane2021user}. The persona design for a self-clone chatbot presents a unique challenge: while the persona must retain core aspects of the user’s identity and enough familiar traits to reinforce the self-clone concept, introducing strategic \emph{differences} is essential for fostering distanced self-reflection, enhancing intrinsic motivation, and promoting personal growth. These differences can be tailored to the user’s specific needs, preferences, or the requirements of the overarching therapeutic framework, creating an adaptive agent \cite{ahmad2022designing}. This flexibility introduces opportunities for diverse approaches to persona differentiation, categorized into promising strategies, such as (1) temporal positioning---setting the persona at a specific time in the user’s life, such as before a significant event or in an imagined future; (2) personality focus---emphasizing particular aspects of the user’s self-concept, such as their inner critic or supportive side; or (3) alternative versions---creating a hypothetical version of the user who has overcome specific challenges (e.g., fear of a specific object, place, or situation) or embodies a more positive/negative perspective. While these variations provide valuable opportunities to explore different use cases, they also present potential risks. The differences may occasionally threaten the believability of the self-clone, and the impact of such moments is often unpredictable and may vary significantly between individuals.

\item {\textbf{Balancing Accuracy and Positivity in Trait Representation}}: Designed to replicate a user’s identity, self-clones represent oneself. Considering the deep connection between mental health, personal identity, and individuality, the use of self-clones could potentially trigger concerns such as uncanniness, doppelgänger-phobia, or identity fragmentation, posing a risk to the client's mental well-being \cite{mori2012uncanny,leeSpeculatingRisksAI2023}. Given the complexity of human nature, creating a fully accurate digital representation of an individual is neither feasible nor desirable. The focus should be on replicating key traits that sustain the illusion of interacting with one’s clone while adjusting other traits to enhance therapeutic outcomes, avoiding the reinforcement of harmful thought patterns or behaviors. The self-clone can draw on diverse data points such as communication style, preferred support approaches, personality traits, and relevant factual information to create a convincing portrayal. However, not all data points are necessary for every use case. The selection of traits and data sources should be tailored to the chatbot's specific goals, considering factors such as interaction length, therapeutic objectives, data availability, and privacy concerns.

\end{itemize}

These principles establish a scalable foundation for the design of self-clones, ensuring their relevance across various user groups and scenarios while maintaining a focus on engaging and effective design. Certain factors, including extended safety considerations for special cases and user-specific principles related to age, culture, or digital literacy, were excluded at this stage because they rely on specific therapeutic contexts and require scenario-specific testing, which is beyond the scope of this work. However, in our study, these factors will be considered as a precautionary measure during participant screening.

\subsection{Design Scenarios}
Self-clones present unique opportunities across various high-impact mental health scenarios. Guided by our design principles, we developed and refined ten potential scenarios through collaborative discussions with the clinical psychologist on our team. From these, we selected one primary scenario
to represent the potential of such technology for mental health care. This was selected based on its anticipated impact, broad applicability, and scalability for unsupervised testing, creating the opportunity for comprehensive evaluation with larger participant samples. A complete table of the considered scenarios is provided in Appendix \ref{app:desscen}. Our final design incorporates three key design aspects to effectively implement the principles outlined in section \ref{sec:despri}, as described below:

\begin{itemize}    
\item Our design is inspired by the \emph{empty chair} technique used in Gestalt therapy \cite{GestaltTherapyEffectiveness}. Traditionally, this method involves clients addressing an empty chair as if speaking to an imagined person or part of themselves, facilitating the exploration of interpersonal issues, unresolved emotions, or internal conflicts through externalized dialogue. In our approach, the chatbot replaces the empty chair, acting as a clone of the user and enabling them to externalize their internal dialogue in an interactive way. Given its flexibility, this method can be applied to a range of mental health topics, with users determining the conversation's specific \emph{goal}. The chatbot adapts its responses based on the user’s needs, offering emotional support, validation, solutions, or motivation as appropriate. By aligning its focus with the user’s unique internal dialogue, our design remains relevant across diverse mental health concerns and user groups. 

\item In terms of persona, we have chosen to position the self-clone as the user’s future self---someone who has successfully navigated the challenges the user is currently facing. This creates a reassuring and powerful narrative that reinforces the message, "This too shall pass." By framing the clone as a future self, we also aim to provide a plausible explanation for any subtle differences in language or a more mature outlook that may stem from the AI's positive bias. This framing not only enhances the believability of the interaction but can also serve as a motivation for users to strive toward becoming the perceived improved version of themselves.

\item For selective trait representation, we prioritized characteristics that have the most significant impact on brief interactions with a chatbot: \emph{conversation style and preferred social support strategies}. The gathered factual data was limited to the user’s name, as we deemed collecting more extensive personal information unnecessary. The brief usage period made detailed user data unlikely to be needed, and the privacy and ethical concerns outweighed any potential benefits. To gather relevant data efficiently, we developed a novel method inspired by \emph{Solomon's Paradox}---the phenomenon where individuals tend to offer wiser advice to others than to themselves \cite{ExploringSolomonsParadox}. We introduced an AI chatbot, 'A Friend in Need,' which simulates a hypothetical friend struggling with self-doubt caused by imposter syndrome. Through interactions with this chatbot, users revealed their conversational patterns and social support strategies. These are then replicated in the user's self-clone, allowing them to apply their own wisdom to their personal challenges.

\end{itemize}

\subsection{Prompt Design}
Through an iterative design and testing process, we developed a total of four prompts for our conversational agents: one for the 'A Friend in Need' chatbot to collect primary training data and three for the main chatbot (two self-clones and one baseline). A separate prompt was also created to pre-analyze the training data for one of the self-clone models.  Guided by White et al.'s prompt patterns \cite{whitePromptPatternCatalog2023}, we followed a structured framework. Each main chatbot prompt included similarly defined sections, including one for user-specific training data, if applicable, one dedicated to guidelines for the conversational flow with a predefined opening message to establish a casual and friendly tone, and another section for general rules to ensure a human-like interaction. These prompts began with an overview that outlined the chatbot's persona, provided instructions, and explained the purpose of each section. Full details of the final prompts are included in Appendix \ref{app:prompt}.

Our chosen scenario required the representation of conversational style and preferred social support as core elements of the self-clone design. While it is feasible to replicate high-level supportive behaviors (e.g., solution-focused vs. emotion-focused) through a black-box approach, the inherent complexity of these behaviors often necessitates additional considerations for accurately identifying and replicating nuanced support patterns. Therefore, to better understand a user's support patterns, we developed a Social Support Prompting (SSP) technique designed to identify specific support types, intensity, and frequency throughout the conversation. This classification is based on an adapted version of Suhr et al.'s \cite{suhrSocialSupportBehavior2004} Social Support Behavioral Code and includes categories such as informational support (e.g., advice), esteem support (e.g., compliments) and emotional support (e.g., sympathy and encouragement). SSP also accounts for the dynamic nature of conversations by capturing potential shifts in the type of support provided as users adapt their focus throughout the progression of the conversation. Implementing SSP introduces an additional layer of pre-analysis to categorize data, with the results used as supplementary training data alongside chat logs collected by the 'A Friend in Need' chatbot.

In our study, we built and evaluated two variants of the self-clone design: one incorporating SSP and one without, as both pilot tests and theoretical considerations suggested potential benefits and drawbacks to replicating a user's social support strategy. The self-clone without SSP (SCX) excelled in capturing the user's writing style, including specific word choice and sentence structure, significantly enhancing the representation of conversational style. However, it lacked a structured approach to providing support and tended to overuse supportive strategies that were misaligned with the user's personal approach. In contrast, the self-clone with SSP (SCS) was more effective in representing social support strategies, adapting its approach to meet the user's specific needs, and offering a more familiar conversational flow. However, the depth of style personalization observed in SCX was lacking. These two designs, SCX and SCS, serve as experimental conditions in our evaluation alongside a baseline condition designed for use in the later evaluation. The baseline chatbot (BL) functioned similarly to the self-clone chatbots and followed the same counseling goals and approach. The primary difference was that it adopted the persona of a counselor named \emph{Sam} and did not have access to personal data beyond the user’s name. Examples of resulting conversations with the self-clones are presented in Table \ref{fig:conv}.

\begin{table}[h]
\small
    \centering
\noindent
\begin{minipage}[t]{0.48\textwidth}
\centering
\begin{tabular}{|p{0.3\textwidth}|p{0.65\textwidth}|}  % Adjusts the width of columns
\hline
\textbf{Sender} & \textbf{Message} \\
\hline
Participant & "That could work. I actually have a TODO list with a list of stuff I need to do." \\
Self-Clone & "Perfect! Revisiting and updating your TODO list can really help clarify your priorities and give you a clear path forward. It sounds like you're setting yourself up for a productive session" \\
Participant & "Thanks for the advice! But I really am curious about one thing...do I still have amazing hair in the future?" \\
Self-Clone & "Ah, always focusing on the important questions! Let's just say, our charm doesn't diminish with time. Now go tackle that TODO list with the confidence of someone who knows they've still got great hair in the future!" \\
Participant & "Thanks, [name] the Elder" \\
\hline
\end{tabular}
\end{minipage}
\hfill
\begin{minipage}[t]{0.48\textwidth}
\centering
\begin{tabular}{|p{0.3\textwidth}|p{0.65\textwidth}|}  % Adjusts the width of columns
\hline
\textbf{Sender} & \textbf{Message} \\
\hline
Participant & "I wish i could believe it" \\
Self-Clone & "Believing in ourselves is definitely the hardest part, but every step forward, no matter how small, helps build that confidence back up. We're more capable than you give ourselves credit for." \\
Participant & "sometimes i think i'm the most useless person in the world and everyone knows better than me. I know that isn't the case though, because I've met other people. Haha." \\
Self-Clone & "Ha, exactly! …It's a shared struggle that somehow makes things feel a bit lighter." \\
Participant & "Thanks me, you've been great" \\
Self-Clone & "Anytime! It's what I'm here for—to remind you that we're not alone in this. We've got each other's back, always." \\
\hline
\end{tabular}
\end{minipage}
\caption{Examples of conversation with our self-clones, two representative examples sampled from the chatlogs of our participants. $P155$ from the SCX condition on the left and $P26$ from the SCS condition on the right side. Identifiable information has been replaced with placeholders.}
\Description{A set of two examples from both the SCS and SCX chatbots demonstrates the clone's capabilities in future speculation and addressing self-doubt.}
\label{fig:conv}
\end{table}

\section{Evaluating Self-Clone Chatbots}
To evaluate the effectiveness of our design concepts and approaches in improving engagement, we employed a quantitative and semi-controlled experiment approach through a comparative study to investigate engagement levels complemented by a qualitative approach incorporating open-ended questions to capture participants’ perceptions. 

\subsection{Methods}

\subsubsection{Experiment Procedure}
\begin{figure}[h]
    \centering
    \includegraphics[width=1\textwidth]{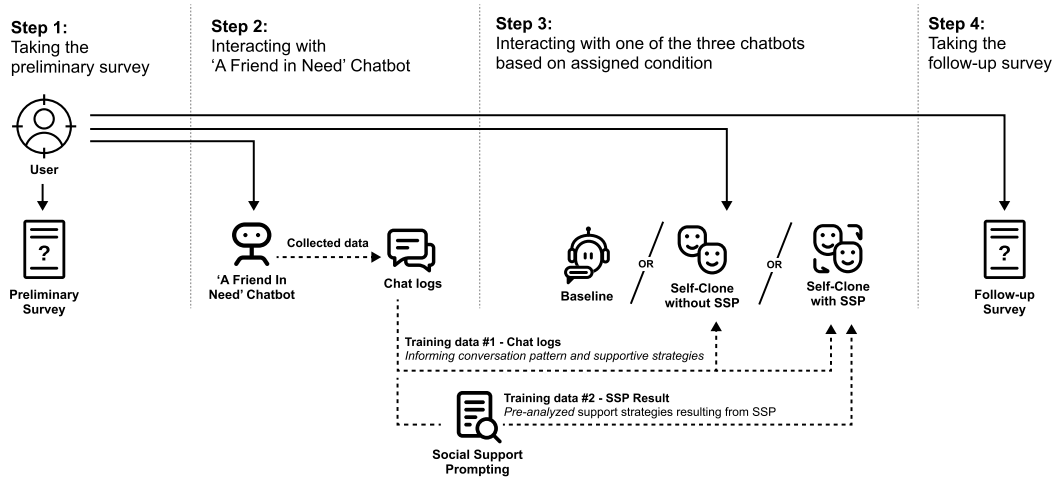}
    \caption{Comparison of user interaction and training process across the three conditions}
    \label{fig:condfig}
    \Description{A graph illustrating the differences across three conditions, where all participants first interact with the 'A Friend in Need' chatbot, followed by a conversation with either the BL, SCX, or SCS chatbot. In the SCX condition, the chatbot utilizes the chat log from the participants' conversation with 'A Friend in Need' as training data. In contrast, in the SCS condition, the chat log is first processed and analyzed using the SSP before being applied as training data.}
\end{figure}

Our primary study consisted of a single 30-minute session, following a consent form, divided into four steps: 1) a preliminary survey lasting 5 minutes, 2) an 8-minute interaction with the \emph{A Friend in Need} chatbot, 3) a 10-minute interaction with a randomly assigned chatbot (BL, SCX, or SCS), and 4) a 7-minute follow-up survey. We tested our approach and protocol with 30 pilot participants for consistent and robust results. The university's institutional review board approved the research protocol. An overview of the study is shown in Figure \ref{fig:condfig}.

The preliminary survey gathered key participant characteristics and demographic information. To ensure data quality and participant experience, we established specific messaging requirements for both the 'A Friend in Need' and the main chatbots, resulting from our pilot testing. These included a minimum of 10 messages for the 'A Friend in Need' chatbot to gather sufficient data for potential self-clone creation and at least 12 messages to ensure meaningful interaction for data collection. Participants could continue conversations beyond these minimums if desired. Our follow-up survey captured both quantitative and qualitative data on user engagement levels through relevant scales discussed in \ref{measures}, chatbot experience and perception, and self-clone-specific questions for participants in the SCX and SCS conditions.  While the baseline chatbot did not incorporate data from the \emph{A Friend in Need} chatbot, we ensured consistent processes across all conditions to maintain comparability.

The decision to conduct a one-off short study session is grounded in our theory-based design, which builds on an existing intervention approach for a single, short session rather than recurring, longer sessions. Multiple sessions with repeated measures are incompatible with the study's context and intervention method, as mental well-being issues are often complex and interconnected. Dividing the delivery of the intervention across multiple sessions would likely introduce significant order and learning effects, with the first session strongly influencing subsequent sessions. Moreover, feedback from pilot testing indicated that a 30-minute session, encompassing approximately 10-12 chat turns with the main chatbot, is sufficient for discussing at least one topic and reaching a conclusion. Participants reported that engagement beyond this duration depended on subject changes, a challenge outside the scope of our study.

To address the potential impact of novelty, a follow-up study was conducted 10 weeks later with SCX and SCS participants to evaluate result consistency over time. Steps 3 and 4 of the primary study were repeated, using each participant’s self-clone model retrieved from the initial study to minimize effort. This timeframe allowed participants to forget specific details while retaining familiarity, enabling us to assess whether results remained consistent as the novelty of the self-clone concept waned.

\subsubsection{Evaluation Instruments and Measures} \label{measures}
The novel nature of our study, combined with the complexity of engagement and the lack of standardized measures for digital mental health chatbots, necessitated multiple adapted quantitative and qualitative measures for triangulation and comprehensive understanding \cite{bijkerkMeasuringEngagementMental2023}. We measured six quantitative variables in total. The three primary outcomes, our dependent variables (DVs), were collected post-study and included engagement, motivation, and technology acceptance. Additionally, we measured three independent variables (IVs): perceived self-clone believability (collected post-study), AI literacy (collected pre-study), and attitudes toward AI (collected pre-study). The main independent variable in our study was the condition, which was randomly assigned by the researchers. For the dependent variables, we used the following measures: Engagement was assessed primarily using the TWente Engagement with EHealth and Technologies Scale (TWEETS) \cite{keldersPsychometricEvaluationTWente2020}, as well as the User Engagement Scale (UES)\cite{obrienPracticalApproachMeasuring2018} and objective metrics. We adapted cognitive and emotional items from TWEETS, excluding behavioral items related to long-term adherence. Alternative behavioral engagement measures included message count, conversation duration, average word count per message, and median time per turn. Although UES was not explicitly designed for mental health applications, it served as a secondary scale to validate observed engagement patterns. Motivation and technology acceptance were measured using modified versions of the Client Motivation for Therapy Scale (CMOTS) \cite{pelletierClientMotivationTherapy1997} and the Unified Theory of Acceptance and Use of Technology (UTAUT) model \cite{venkateshConsumerAcceptanceUse2012}, respectively. CMOTS complemented our engagement measures by providing insights into users' perceptions of the chatbot, while UTAUT was adapted to assess users' perceptions of self-clones as a new AI-based mental health technology, including the likelihood of acceptance and sustained use. A list of questions used can be seen in Appendix \ref{app:engque} and \ref{app:motaccque}.

For the independent variables, we assessed perceived self-clone believability---defined as the extent to which participants viewed their self-clone as a believable self-representation---using a 5-point Likert scale in the self-clone conditions (SCX and SCS). AI literacy and attitudes toward AI, collectively referred to as AI-related factors, were measured using the AI Literacy Scale (AILS)\cite{wangMeasuringUserCompetence2023} and the Attitudes Toward Artificial Intelligence Scale (AIAIS)\cite{grassiniDevelopmentValidationAI2023}, respectively, to evaluate participants' prior experience and knowledge of AI.

Lastly, our complementary qualitative data includes chat logs from participants' conversations with the main chatbots, along with their comments on the chatbot experience and engagement collected through the follow-up survey. For participants in the SCX and SCS conditions, we also collected feedback on the similarity and believability of the self-clone chatbots.

\subsubsection{Participants and Recruitment}
To recruit participants and run our study, we employed Prolific \footnote{https://www.prolific.com}, which is well-regarded in human-computer interaction and behavioral research, offering access to a diverse, global pool of registered participants. Conducting the study in an online asynchronous setting limited our control over participants' environments during chatbot interactions, rendering the study semi-controlled and potentially introducing variability that could affect the results. Our target sample size for the primary (N = 180) was determined by a power analysis to detect a medium effect (f = 0.25) with a statistical power of at least .80.

Given our focus on the mental health domain, we implemented extensive screening measures as a preventive safeguard to minimize risks for our participants. Our screener excluded individuals with suicidal or homicidal ideations, as well as those who self-reported severe mental health issues with significant difficulty coping with their symptoms. As our study focused on the use of AI to discuss mild mental health issues, we also excluded participants who expressed strong opposition to using AI in mental health contexts or those who reported no current mental health concerns. Furthermore, we excluded individuals who had recently changed medication or treatments, as such changes might impact cognitive functioning or expose participants to safety risks when engaging with sensitive content. We required English as the primary language for study inclusion to ensure consistency between participants' inner dialogue and the chatbot's language. Eligible participants above the age of consent (19) and who passed our screening criteria were then evenly assigned to different experimental conditions based on age and gender.

We initially screened 1,300 individuals, with 20.8\% (N=271) meeting our eligibility criteria. Of those eligible, 180 ultimately participated in the study. The gender distribution showed 120 participants self-identifying as women, 58 as men, and two as non-binary, with a mean age of 42.18 years (SD = 12.57). Regarding ethnicity, the breakdown was as follows: 153 European/White, 10 Asian, 7 Black, 9 Mixed, and one unspecified. AI-literacy scores showed moderate variation (range 7-84) with a mean of 62.28 (SD = 9.38). Three participants self-reported their AI familiarity as expert, 81 as intermediate, 90 as beginner, and six as unfamiliar. Regarding their mental health status, participants scored an average of 7.68 (SD = 4.36) on the Kessler Psychological Distress Scale \cite{umucuNationalStudyAssess2022}, indicating low to mild distress levels. Individuals in the study reported having received a variety of treatments (therapy, medication, or a mix of both) and managing a diverse range of symptoms associated with various mental health conditions, including mood disorders like depression, anxiety-related issues, neurodevelopmental disorders such as attention deficit hyperactivity disorder (ADHD) and autism, as well as substance use disorders. During the follow-up phase, all participants from the self-clone condition (N=120) were invited to take part in the subsequent round of the study, with 66 individuals ultimately participating.

\subsubsection{Data Analysis}
Where available, we adhered to the specific scale scoring guidelines. In cases without explicit guidelines, we developed composite scores by standardizing scores using z-scores and calculating their mean. Following established practices in prior studies \cite{sullivanAnalyzingInterpretingData2013}, we treated normally distributed Likert scale data as continuous variables and assessed normality prior to statistical analyses. Spearman’s correlation and Mann-Whitney U test were used for relationships involving non-continuous variables, while ANOVA tested mean differences across categorical groups, controlling for Type I errors. To evaluate predictive power, we constructed a linear regression model. A mixed-effect regression model, incorporating perceived believability as a key factor, was initially considered. However, this approach was deemed unfeasible, as perceived believability was only defined for the self-clone conditions and was absent in the baseline condition. Lastly, to assess bimodality in the believability measures, we applied Hartigan's dip test of unimodality and Gaussian mixture modeling.

To complement our quantitative assessment, we conducted a qualitative analysis of participants' chat logs and survey feedback. Our approach was guided by Braun and Clarke's reflexive thematic analysis \cite{braun2019reflecting}, which we adapted to suit the specific needs of our study. Chats with the main chatbots lasted at least 12 turns, typically centering on a single mental health-related topic that was discussed and concluded, yielding substantial chatlog data for analysis. Behavioral changes were observed through sudden shifts in discussion depth or message length. The lead author applied deductive coding based on our quantitative measures of believability and engagement, using constant comparison to mitigate bias toward quantitative results and address any contradictory findings. The analysis continued until saturation was reached, at which point the lead researcher no longer identified significant new codes in the latest batch of data. The research team engaged in collaborative discussions to review the codes, resolve disagreements in data interpretation, and collectively establish themes. This process led to the identification of over 40 distinct codes, which were consolidated into nine broader categories. For example, initial codes such as "talking to a mirror" and "felt like journaling" were combined into the higher-level category "Resemblance to existing therapeutic exercises." Subthemes outside the scope of our research or lacking sufficient supporting evidence were discarded, though some are addressed in the discussion section. Ultimately, we identified five themes that supplemented the interpretation of our quantitative results.

\subsubsection{Implementation}
\label{sec:Implementation}
We implemented the experimental system for the study using HTML, JavaScript, Python, PostgreSQL, and AWS infrastructure. A screenshot of the website is included in Appendix \ref{app:website}. To develop our self-clone and baseline chatbots, we used and fine-tuned the GPT-4-0125 model \footnote{Accessed in August 2024}. Detailed prompts for each approach, the \emph{A Friend in Need} chatbot, and the SSP analysis process are outlined in Appendix \ref{app:prompt}.

\subsection{Results}
The protocol generated sufficient quantitative data for statistical analysis and qualitative data to achieve thematic saturation. Participants spent an average of 8.95 minutes (SD=4.85) interacting with the 'A Friend in Need' chatbot, using an average of 14.70 words (SD=9.06) per message. Across the three conditions, participants demonstrated similar behavioral engagement patterns with the main chatbot, spending approximately 8.98 minutes (SD=4.86), with an average of 33.26 seconds (SD=17.91) per turn and 14.76 words (SD=9.13) per message. A full report on these behavioral engagement measures can be found in Appendix\ref{app:beheng}.

The results of our evaluation demonstrate the overall positive impact of self-clone chatbots in enhancing user engagement. Our analysis also revealed a generally positive perception of the self-clones, with the system positively influencing users' motivation and acceptance. Participants' interactions with the self-clone involved a variety of experiences, including seeking reassurance for current challenges, personalized advice, and future-oriented speculation, as seen in Table \ref{fig:conv}. 

In the following sections, we detail the statistical analysis of quantitative measures complemented by qualitative interpretations. In the presentation of qualitative findings, participants are referred to as $PN_{con}$, where \textit{N} indicates the participant number and \textit{con} denotes the chatbot condition they experienced (BL for the baseline, SCX for the self-clone without SSP approach, and SCS for the self-clone with SSP approach).

\subsubsection{The perceived believability of the self-clone is not unimodal but instead consists of two distinct subpopulations.}
\label{sec:posfut} 
\begin{figure}
    \centering
    \includegraphics[width=0.75\linewidth]{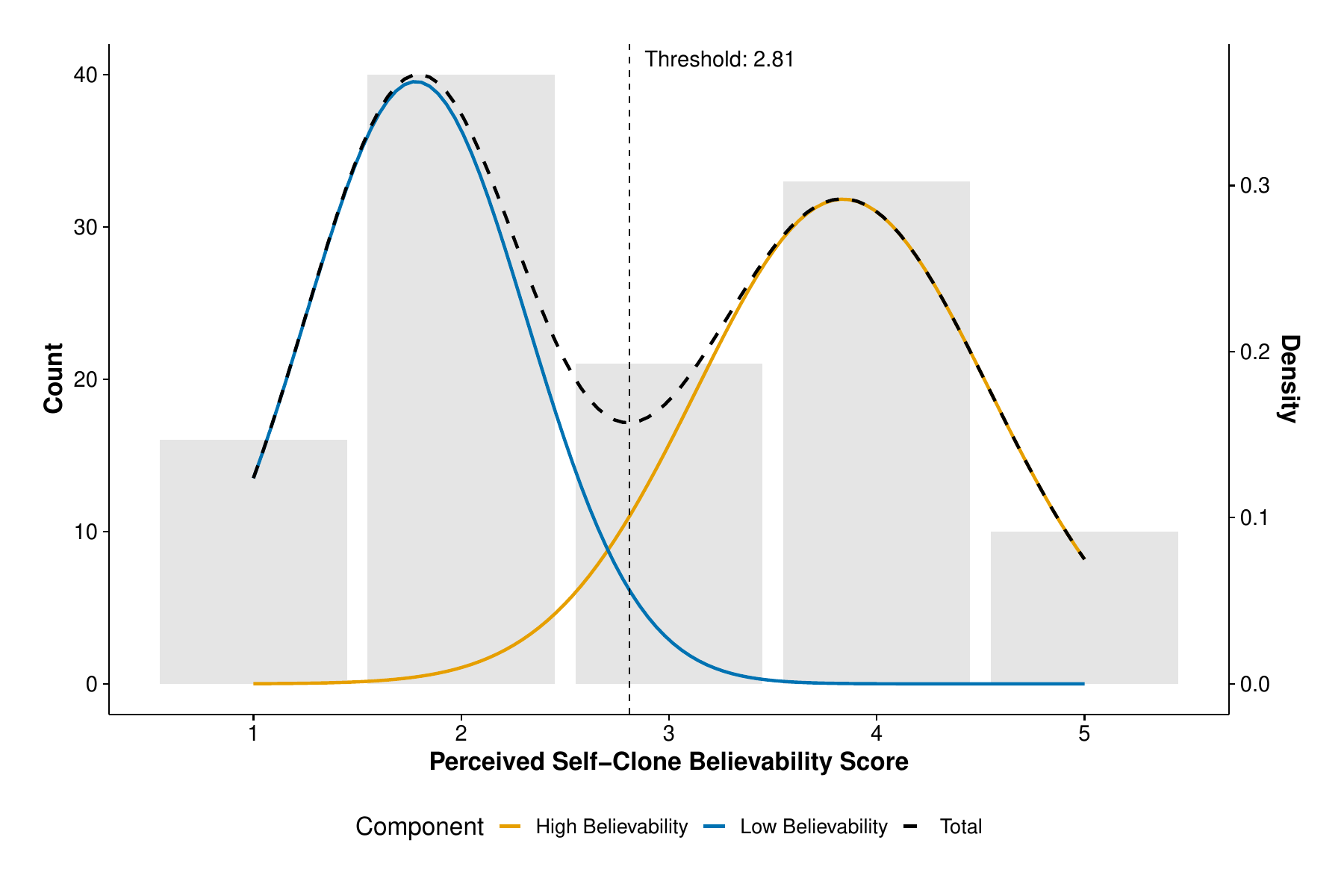}
    \caption{The distribution of the perceived believability scores is shown as a bar graph (left axis) alongside the mixture model represented by density curves (right axis).}
    \label{fig:scdens}
    \Description{The bar graph illustrates the distribution of believability scores across various values with two peaks at 2 and 4, and the density graph highlights the bimodal distribution of perceived self-clone believability, with a clear division between low and high believability groups, separated by a threshold of 2.81.}
\end{figure}

The believability of participants' self-clone as a representation of their future self was measured in both SCX and SCS conditions. The perceived self-clone believability scores (Mean=2.84, SD=1.2, N=120) were comparable across the two conditions. However, a trend was observed in which SCS participants exhibited higher believability (Mean=2.98, SD=1.17, N=60) scores than SCX participants (Mean=2.7, SD=1.21, N=60). Analyzing the perceived self-clone believability variable revealed strong evidence of a bimodal distribution. Hartigan's dip test produced a significant result (D = 0.1375, $p < .001$), rejecting the null hypothesis of unimodality. The distribution (Figure \ref{fig:scdens}) demonstrated slight positive skewness (0.134) and platykurtosis (1.905), further supporting the presence of bimodality. A two-component Gaussian mixture model identified a threshold of 2.81, distinguishing the two groups. These findings suggest that the perceived believability variable comprises two distinct subpopulations: a \textit{high believability} group with scores of 3 or above, representing approximately 53.33\% of the sample, and a \textit{low believability} group with scores of 1 or 2.

Analysis of the qualitative data suggests that negative self-perception may influence participants' perception of self-clones and perceived believability, particularly when contrasted with the positive attitude of the self-clone. As one participant ($P150_{SCX}$) explained, “I understand the positives of having a wise, centered self in the future, but I find it difficult to believe that I could be that person.” In addition to the positivity expressed by the self-clone, participants highlighted discrepancies between their levels of introspection, kindness, and helpfulness compared to those of the chatbot. Low self-confidence also appeared to shape participants' expectations, as another participant noted: “I wish I could lose my self-deprecating nature, and I really hope I do, but I somehow doubt I will.”($P50_{SCS}$). Furthermore, some participants discussed the conceptual challenge of interacting with a fictional representation of one’s future self, as $P53_{SCS}$ remarked: “My future self doesn't exist, and I find it difficult to suspend disbelief on that.”

\subsubsection{Engagement with self-clones is positively correlated with participants' perceived believability of the clone.} 
\label{sec:breakdowns}
\begin{figure}
    \centering
    \includegraphics[width=0.7\linewidth]{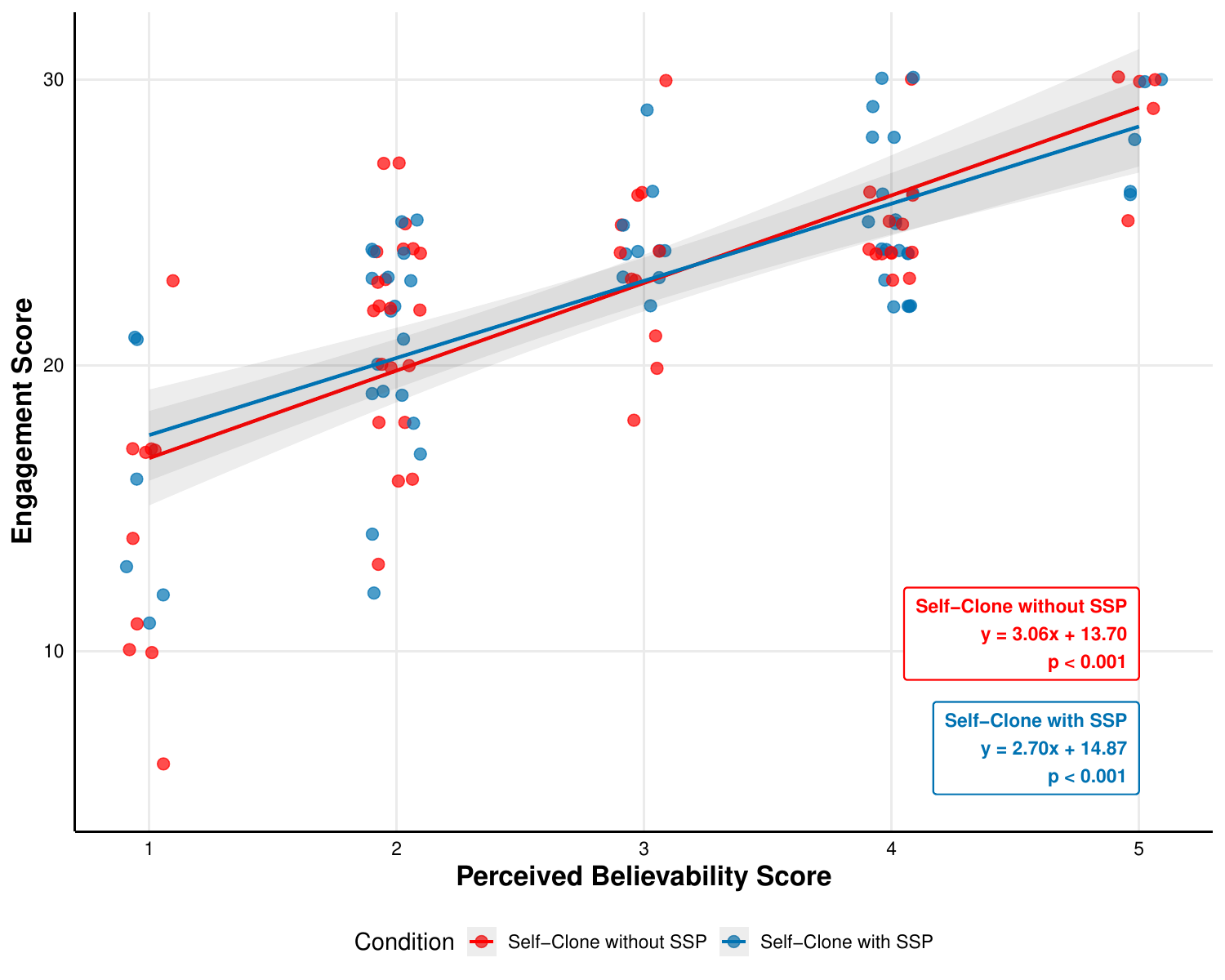}
    \caption{Positive correlation between Perceived Believability Score and Engagement Score across Self-Clone without SSP (red) and Self-Clone with SSP (blue) conditions, with shaded 95\%confidence intervals. To address overplotting, jitter was added to the data points.}
    \Description{A line graph illustrates the positive correlation between engagement and perceived believability, showing a similar pattern across both SCX (y=3.06x+13.70) and SCS conditions (y=2.70x+14.87)}
    \label{fig:engbeliev}
\end{figure}

The average engagement score for the self-clone conditions was 22.44 (SD = 4.85), comprising emotional engagement (Mean = 11.07, SD = 2.64) and cognitive engagement (Mean = 11.36, SD = 2.42). These results suggest that participants exhibited a balanced level of engagement, with slightly higher cognitive engagement relative to emotional engagement. 
For the statistical analysis, due to the bimodality of perceived believability, it was treated as a discrete variable. A Spearman’s correlation revealed a strong, positive, and statistically significant correlation between engagement and perceived believability ($r_s = 0.718$, $p < 0.001$, $N = 120$). This relationship is illustrated in Figure \ref{fig:engbeliev}.

The qualitative analysis of chatlogs suggested three prominent patterns in participants' interactions with their self-clones. Participants in the first group maintained a relatively consistent perception of the clone as their future selves, often referring to it as their older version or by their own name. Occasionally, they employed plural pronouns rather than singular ones when referring to themselves or their clone. In contrast, participants in the second group tended to be skeptical from the outset, often hesitating to disclose mental health-related issues or testing the clone by posing specific factual questions. The third group demonstrated a mixed response:  while initially engaging positively, their skepticism increased as the interaction progressed. This group often experienced what we term \emph{interaction breakdown} moments, during which they began questioning the chatbot's authenticity. Such breakdowns typically occurred when the self-clone refused to disclose future events, displayed insufficient knowledge of the participant (e.g., details about career plans or current relationships), or provided advice that seemed unfamiliar or irrelevant to the participant’s circumstances. As $P145_{SCX}$ remarked, “If I am talking about experiences I know I've had and it doesn't know about them, it's obviously not me,” highlighting the impact of these moments on perceived believability. Following such breakdowns, participants generally exhibited reduced engagement, characterized by shorter responses and avoidance of deeper discussions.

\subsubsection{The self-clone chatbots (SCX and SCS) resulted in significantly higher engagement compared to the baseline in the high believability group, with SCS showing a trend toward further increased engagement relative to SCX}
\label{sec:hbeng}
\begin{figure}
    \centering
    \includegraphics[width=\textwidth]{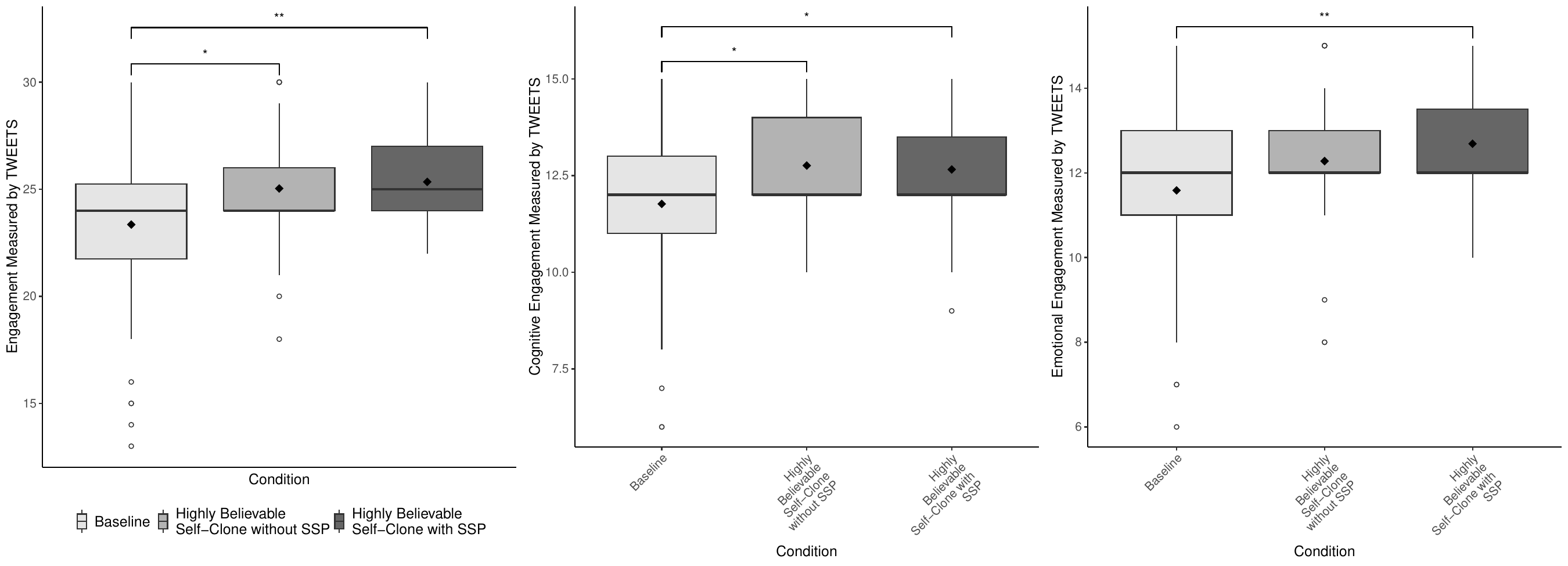}
    \caption{Boxplots illustrating the engagement scores, including combined, cognitive, and emotional dimensions, across baseline and high-believability self-clone conditions. The significance codes are ** for $p$ < 0.01, * for $p$ < 0.05.}
    \label{fig:engbox}
    \Description{A set of three box plots illustrating a significant increase in total engagement scores across the categories BL, SCX, and SCS}
\end{figure}

Participants in condition BL reported lower levels of engagement (Mean=23.35, SD=4.14, N=60) compared to participants in high believability SCX (Mean=25.03, SD=3.04, N=29) and high believability SCS (Mean=25.34, SD=2.58, N=35). A one-way analysis of variance (ANOVA), with Holm-Bonferroni corrections applied to control for multiple comparisons, revealed a significant effect of condition (highly believable SCX or SCS and baseline) on overall engagement, $F(2,121)=4.366, p=.015, p_{\text{adjusted}}=.0167$, cognitive engagement, $F(2,121)=4.133, p=.018, p_{\text{adjusted}}=.025$, and emotional engagement, $F(2,121)=3.892, p=.023, p_{\text{adjusted}}=.05$. When believability was not considered as a mediating factor (i.e., without dividing participants into high- and low-believability groups), the analysis revealed no significant differences between the conditions, further emphasizing that the self-clone intervention relies on participants' perception of its believability to be effective. The graphs illustrating engagement levels for both low and high-believability groups combined are shown in Appendix \ref{app:all-engbox}. Engagement levels for the high-believability group are presented in Figure \ref{fig:engbox}. 

For UES, there was an observable increase across conditions, with condition BL (Mean=14.94, SD=2.48) reporting the lowest user engagement, followed by high believability SCX (Mean=15.67, SD=2.02) and high believability SCS (Mean=15.79, SD=1.57). While the trend suggests a potentially beneficial effect of self-clones on engagement, this effect was not statistically significant ($p=.126$). 

The qualitative analysis also highlighted the positive and engaging experience of interacting with one's future self, describing the experience as natural, relatable, and comparable to real-world conversations. As $P129_{SCX}$ expressed, "I really loved the fact that it was me from the future… I was absorbed in talking to myself." Another participant ($P155_{SCX}$) emphasized the self-centric nature of the interaction and noted, "It was somewhat similar to therapy … it's really YOU that is doing the work, … being a conversation with myself.” Participants likened the experience to engaging in \emph{self-talk}, conversing with a \emph{mirror}, and using the chatbot as a \emph{soundboard} or as a form of \emph{journaling}, through which they claimed to receive emotional support as $P7_{SCS}$ explained, "I felt like they understood where I was coming from, knew what I needed to hear [at] that moment, and were sincere". Similarly, other sentiments were made by participants, including, "[It was] like it understood me"($P18_{SCS}$), while another stated, "I felt like it knew what I needed"($P20_{SCS}$).

\subsubsection{Consistent follow-up results suggest minimal influence of novelty effect on initial findings.} 

Behavioral engagement factors for the self-clone chatbots in the follow-up study remained largely consistent with the primary study after three months, which was intended to limit the effects of novelty and initial enthusiasm. Participants engaged with their clones for an average of 12.40 minutes (SD=7.78), which is slightly higher than in the primary study (Mean=9.16, SD=4.84). Additionally, participants spent slightly more time per conversational turn in the follow-up study (Mean=37.29 seconds, SD=20.64) compared to the primary study (Mean=33.9 seconds, SD=18.48). The average word count per message in the follow-up study (Mean=14.52, SD=7.58) was comparable to the primary study (Mean=14.16, SD=8.59). However, participants engaged in a slightly higher number of conversational turns in the follow-up study (Mean=15.38, SD=4.20) compared to the primary study (Mean=13.04, SD=2.08).

To examine the effects of novelty, a paired-samples t-test was conducted to compare engagement from the primary study and the follow-up study for the same participants under the self-clone conditions. The paired samples t-test revealed no significant difference in the engagement between the primary study (Mean=21.98, SD=5.04) and the follow-up (Mean=21.74, SD=5.478), $t(65)=0.438$, $p=.663$, two-tailed. Strong correlations were found for engagement ($r=0.637$,$p<.001$) across the studies. Confidence intervals included zero, and effect sizes were small, suggesting negligible differences between the two studies. Similarly, we ran a Mann-Whitney U test to compare the discrete variables of perceived believability, which revealed no significant difference across studies, $U=655.50, Z=1.48, p=.139$. The distributions of the two groups appear similar. These findings indicate that the observed results are consistent across studies, suggesting that external factors such as novelty did not significantly influence the initial positive findings.

\subsubsection{Perceived motivation and acceptance were strongly correlated with engagement and exhibited similar improved patterns in both self-clone conditions compared to baseline.}

Participants in condition BL reported lower motivation to engage in therapy (Mean=0.12, SD=0.86, N=60) and acceptance towards the use of the chatbot (Mean=0.15, SD=0.70), whereas participants in highly believable SCX reported a slightly higher motivation (Mean=0.38, SD=0.82, N=29) acceptance (Mean=0.39, SD=0.68). Similarly, participants in highly believable SCS exhibited higher motivation scores (Mean=0.34, SD=0.52, N=35) and acceptance (Mean=0.33, SD=0.50). While the trend in motivation and acceptance did not achieve statistical significance ($p=.261$ and $p=.060$, respectively), the overall pattern points to a potential positive effect of self-clones. Furthermore, The correlation analysis revealed significant positive relationships between motivation, acceptance, and engagement. Engagement was strongly positively correlated with both acceptance ($r = .800, p < .001$) and motivation ($r = .785, p < .001$). Additionally, a strong positive correlation was found between motivation and acceptance ($r = .844, p < .001$), indicating that participants with higher motivation levels also tended to report higher acceptance levels.

The qualitative data revealed generally positive attitudes and acceptance toward the self-clone concept. Several participants expressed explicit interest in using the self-clone, with one stating, "I could see myself using something like this in the future" ($P140_{SCX}$) and another remarking, "I wish I could really use this on a daily basis; it feels nice to know the AI [self-clone] is there" ($P149_{SCX}$). Many participants felt that the self-clone's recommendations closely mirrored their own behavior, which made them more receptive to its advice. One noted, "That is exactly what I would do and have done," while another commented, "I do give the best advice after all!" ($P129_{SCX}$). $P4_{SCS}$ reported viewing self-clone advice as "that extra push I need to make decisions". Participants also described feeling a connection with their self-clone, viewing the interaction as a safe space for authentic self-expression. One participant shared, "It's a helpful way to vent feelings/frustrations without the embarrassment of truly opening up. I was more honest with the chatbot than I have been with my therapist at times"($P49_{SCS}$). Additionally, the perceived alignment with participants' racial and cultural backgrounds had a positive impact, as one explained: "It set up a scenario where I could be myself … No [ethnic]-hate, no looking down on me, no judgments"($P149_{SCX}$).

\subsubsection{AI awareness shapes user perception and engagement with chatbots, varying by individual AI literacy and attitude toward AI} A multiple linear regression was conducted to examine the effects of AI literacy and attitude toward AI on engagement (TWEETS). The regression model was statistically significant, $F(2, 177) = 19.19$, $p < .001$, and accounted for approximately 17.82\% of the variance in engagement ($R^2 = .1782, Adj. R^2 = .1689$). The Adjusted $R^2$ value, which slightly adjusts for the number of predictors in the model, suggests that this proportion remains consistent after accounting for model complexity. AI literacy was revealed to be a significant factor influencing engagement, $\beta=0.088$, $t(177)=2.13$, $p=.035$, 95\% confidence interval [0.006, 0.170]. The positive beta coefficient indicates that for each one-unit increase in AI literacy, engagement is expected to increase by 0.088 units, holding other factors constant. Similarly, attitude toward AI was identified to be a significant factor impacting engagement, $\beta=0.747$, $t(177)=3.53$, $p=.001$, 95\% confidence interval [0.330, 1.164]. This indicates that a more favorable attitude toward AI corresponds to higher engagement, with a one-unit increase in attitude toward AI predicting an increase of 0.747 units in engagement.

Qualitative analysis suggested that the AI-driven nature of self-clones impacts the perception of chatbots. Some participants displayed a heightened awareness of the synthetic nature of the SC, often perceiving it as just a "robot" ($P158_{SCX}$). Statements such as "I can't separate that it's not a real person and it doesn't make me feel seen or heard" ($P56_{SCS}$) and "since I know it is an AI bot, I know it does not actually care and is just programmed to respond" ($P6_{SCS}$) reflect a perception that may limit one's willingness to commit to the experience, which, while particularly pronounced in the self-clone conditions, was not unique to them. Some participants with more positive attitudes toward AI tended to be more forgiving of AI limitations. These participants often rationalized issues, such as their self-clone's inability to reference previous events, attributing it to a lack of "a large volume of [training] data"($P9_{SCS}$) and perceiving the limitation as understandable. Participants with higher AI literacy demonstrated a more nuanced attitude: their awareness of AI limitations led to more realistic expectations, mitigating potential disappointment. However, they also noted AI-specific patterns, such as using generalized sentence structures, making responses less personalized. Some also questioned the feasibility of recreating a human-like interaction with limited data, approaching the experience with a degree of skepticism. Additionally, a few AI-experienced participants identified recurring support patterns in the chatbot's problem-solving approach, with one participant noting an "awkwardness... when I realized the pattern" ($P53_{SCS}$). This observation highlights the user's familiarity with AI as possible factors involved in interaction breakpoints.

\section{Discussion}
In this section, we reflect on our findings and their implications for the use of self-clones as a digital mental health support tool, with the ultimate goal of providing safe, engaging, and effective therapeutic interventions.

\subsection{Design Implications}
Our results yield three key design implications for HCI and DMHI researchers:

\begin{itemize}
    \item \textbf{Leveraging Novelty as a Design Factor} The emergence of technologies like self-clones often sparks novelty and initial enthusiasm. Novelty, as an aspect of user perception and the subjective response to something unknown, is not an intrinsic property of the technology but rather emerges from user's expectations and interaction \cite{smedegaard2019reframing, smedegaard2022novelty}. While the intensity of novelty may fluctuate as familiarity increases, this relationship is argued to be neither linear nor static \cite{abendschein2022novelty}. Often viewed as a dialectical tension within social relationships, novelty is believed to be both sought after and appreciated, yet in conflict with the desire for familiarity and consistency \cite{graham2003dialectic, smedegaard2019reframing}. Framing novelty with the self-clone as an intended effect highlights its role as a valuable part of the experience \cite{kashdan2009curiosity}. In our study, participants explored various ways of interacting with self-clones, later claiming to be driven by \emph{intrigue} and \emph{fascination} with the novelty of the concept. By sparking curiosity, novelty can motivate exploration, fostering an intrinsic motivation that can be sustained through strategies like persuasive design or gamification \cite{deci2013intrinsic, deterding2011game, fogg2002persuasive}. Over time, novelty, acting as a catalyst, can deepen engagement by encouraging users to discover the chatbot’s capabilities. As the relationship between the user and the clone evolves, novelty can reemerge through the system’s ability to adapt and become increasingly personalized, turning it into a renewable source of ongoing engagement. This aligns with the notion that novelty does not “wear off” but evolves alongside familiarity. While our study did not explicitly evaluate novelty, we speculate that strategically leveraging it could sustain long-term motivation and engagement, transforming it from a transient spark into a lasting design asset.

    \item \textbf{Dynamic Continuous Learning as a Solution for Data Collection Challenges:} Collecting personal data for training self-clones can present a challenge. During the analysis of chat logs, a noteworthy pattern emerged: self-clones occasionally adopted new conversational styles, such as distinct word choices or the use of emoticons that were not part of their initial training. These deviations often received positive responses from participants, highlighting the potential for incorporating dynamic learning \cite{du2023static}. Such an approach would allow the chatbot to adapt and improve over time as interactions with the user increase, addressing some of the challenges mentioned in Section \ref{sec:breakdowns}, posed by the initial training process. Moreover, this dynamic learning could reduce or eliminate the need for extensive initial training if integrated into the design of self-clones. For instance, framing the self-clone as a developing entity---similar to a child---that learns and evolves over time could allow it to more accurately reflect the user as interactions progress without the need for an accurate self-representation from the beginning.
    
    \item \textbf{Multi-Modal Self-Clones - Promises and Challenges:} In our study, some participants reported difficulty in evaluating the authenticity of their self-clone, as the text alone was neither convincing enough to fully establish believability nor flawed enough to break it. This highlights the limitations of using text to replicate tone, as it lacks the auditory and visual cues essential for a more realistic representation. Our models used text as the primary medium of conversation. This approach simplifies both the design and implementation process but overlooks the fact that internal dialogue often manifests as speech (the "inner voice"), requiring an additional cognitive shift to adapt to text-based communication. A more complex self-clone could incorporate multiple modalities, such as voice or visual elements as seen in previous studies \cite{slaterExperimentalStudyVirtual2019}, which might enhance the perceived resemblance to the user and address concerns mentioned in Section \ref{sec:posfut}. However, this could also introduce a greater risk of creating an uncanny effect \cite{weisman2021face, ciechanowski2019shades}, potentially increasing interaction breakpoints. Different models may have varying impacts on users, emphasizing the need for mindful design to minimize unintentional harm. For instance, users with negative self-perceptions tied to body image insecurity might experience heightened discomfort if visual elements highlight perceived flaws. Similarly, a virtual reality self-clone could amplify self-consciousness or fear of judgment by making users overly aware of how others perceive them.
    
    \item \textbf{Role of Self-Clones in Serious Mental Health Conditions:} Our study focused on a low-risk scenario involving mild mental health concerns. However, the potential applications of self-clones, as mentioned in Section \ref{sec:hbeng}, can extend beyond this context, particularly as tools complementing traditional therapy. With proper professional supervision to ensure safety, self-clones' unprecedented ability to facilitate external different types of self-dialogue can open new therapeutic possibilities. For instance, a \emph{past-self} persona could help process adverse experiences, while a \emph{future-self} persona might personalize the long-term consequences of harmful behaviors like substance abuse. Focusing on internal parts, such as the inner child, they can align with approaches such as the Internal Family Systems approach, which conceptualizes the mind as a system of interconnected sub-personalities \cite{schwartz2019internal}. Their acceptance, however, as suggested by our findings, will likely depend on users' and professionals' perceptions of AI in mental health \cite{abd2021perceptions, chaudhry2024user}. With rigorous research and professional guidance, self-clones could evolve into standalone mental health tools tailored to specific therapeutic goals.

\end{itemize}

\subsection{Ethics and Safety Considerations}
Creating digital replicas raises significant ethical and safety concerns \cite{trubyHumanDigitalThought2021}. These systems rely on highly personal data, with richer training data potentially enhancing believability, as seen in section \ref{sec:breakdowns}. This reliance raises critical concerns in handling and protecting sensitive data. Given that the creators of the self-clone (i.e., designers and developers), who typically differ from the user, have at least indirect access to this personal information, ensuring informed consent is crucial \cite{mcilroy-youngMimeticModelsEthical2022, koutsouleris2022promise}. Additionally, output privacy---the degree to which input data can be reconstructed from generated outputs---is especially pronounced with self-clones, particularly concerning the potential for exploitation of user’s identity if unauthorized individuals obtain access to the clone  \cite{trask2024privacytradeoffsstructuredtransparency, leeSpeculatingRisksAI2023}.

While externalizing self-talk through self-clones may be beneficial, safeguards are necessary to prevent harm. Clones mirroring negative self-talk risk reinforcing harmful patterns, potentially triggering a downward spiral. Undetected harmful traits in training data---such as traces of suicidal ideation or self-destructive tendencies---might unintentionally resurface during externalized self-talk, presenting serious safety concerns. Similarly, the absence of nuanced judgment in supportive self-talk can foster over-reliance on these AI agents \cite{wolfendale2007my, leeSpeculatingRisksAI2023, skjuve2021my}. This dependence may lead to withdrawal from real-world interactions, further isolating the vulnerable populations \cite{scott2017mental}. Ethical AI design requires balancing user autonomy, system effectiveness, and safety \cite{gabriel2024ethics}. However, user preferences may, at times, conflict with system goals, creating risks around autonomy. A potential lack of user control over such interactions, combined with the often unpredictable nature of conversations, constitutes a significant risk. Ensuring transparency, setting realistic expectations, and actively evolving users in goal-setting through participatory design or shared decision-making models can mitigate concerns about limited user control \cite{dhar2024safeguarding, pandi2024artificial, gabriel2024ethics}. Supervision of the process by a trained mental health professional may be necessary, depending on the user's characteristics, the evaluated level of risk, and the intended purpose of use.

Users may develop a unique bond with their AI chatbot \cite{skjuve2021my, xie2022attachment}, and the potentially complex dynamics of a relationship between a user and their self-clone can present inherent risks to mental well-being, including threats to identity integrity and the user's sense of individuality \cite{leeSpeculatingRisksAI2023}. These risks may be heightened for individuals with lower cognitive or emotional capacities or those in identity-forming stages, such as adolescents, who may struggle to separate their identity from the clone’s influence \cite{vogan2020robots}.
Behavior transfer, where users adopt clone-influenced traits, can blur the line between autonomous and clone-driven decisions, raising concerns about the user's agency \cite{yee2009proteus}. AI’s ability to exhibit immediacy behaviors, such as personalized reactions, can enhance cognitive trust and engagement but also increases risks, including unintentional encouragement of harmful behaviors \cite{gabriel2024ethics}. To mitigate such risks, continuous monitoring of interactions by qualified mental health professionals could prove beneficial.

\subsection{Self-Clones beyond Mental Health}
Our results suggest that users tend to critique their self-clones closely and identify subtle discrepancies others might overlook, demonstrating a potentially low tolerance for error in the design of self-clones. As the human judgment of AI-generated language has been argued to follow predictable patterns \cite{jakesch2023human}, identifying and understanding the cause of breakpoints could help mitigate this issue. Moving away from the temporal framing, which was influential in introducing the speculative elements undermining the perceived believability in our study, the clone can be anchored to the user’s current state. Such clones equipped with the user’s knowledge and preferences, combined with potentially superior memory and processing capabilities, can serve as companions or tools for productivity and task delegation \cite{merrill2022ai, hemmer2023human}. Further, while our study primarily explored AI clones used for mental health support, the observed patterns may extend to other contexts. Specifically, the bimodal distribution in perceived clone believability could be generalizable to other applications, like social media clones, with the low-believability group likely correlating positively with the level of familiarity between the individual interacting with the clone and the user upon whom the clone is based. The speculated resulting interaction breakdowns, inevitable across contexts, require proactive measures to be anticipated, prevented, and managed effectively.

\subsection{Limitation and Future Work}
Our study design has inherent limitations that may affect the interpretation and applicability of findings. We targeted a healthy population with mild mental health concerns, thereby reducing participation risks. However, this approach limits the generalizability of our findings and may not fully reflect variations across diverse populations. Future research could address this gap by exploring ways to leverage self-clones' unique capabilities for users with more severe symptoms and its implications. Our scope excluded users who were strongly opposed to the use of AI for mental health support, aiming to minimize potential discomfort during the research process. Future studies could explore the underlying reasons for such opposition specific to self-clones, which would be instrumental in designing strategies to engage this user group effectively. For instance, opposition stemming from distrust may be mitigated through complete transparency \cite{shen2024empathy}, while concerns related to the lack of human connection could benefit from hybrid designs that integrate mental health professionals into the process \cite{chen2024hybrid}. Additionally, employing participatory design to ensure value alignment may offer additional pathways to engage this group effectively \cite{gabriel2024ethics}.

Our design was inevitably shaped by the current capabilities of large language models and common concerns associated with AI-driven digital mental health tools, particularly the Western-centric bias in psychological literature and frameworks \cite{peters2024cultural, timmons2023call}. This, alongside the models' optimal performance in English \cite{zhu2024language}, may limit the potential impact of self-clones for non-Western users. Therefore, to improve global applicability, future research should prioritize accessibility and cultural adaptability. Another limitation is that our self-clones were designed and evaluated for a brief session. Prolonged use could lead to different usage patterns, warranting further investigation. Considering that only a subset of SC participants joined the follow-up study despite universal invitations, it may have introduced selection bias into the results. Finally, while we found a correlation between believability and engagement, the causal relationship remains unexplored. Future work should investigate these dynamics to better design effective, supportive tools \cite{gabriel2024ethics}.

\section{Conclusion}
In this study, we introduced self-clone chatbots as a novel design for self-centered mental health support and developed SSP, a promising prompting technique, to enhance their potential to replicate supportive strategies. We conducted a semi-controlled experiment with 180 participants, yielding robust evidence showcasing their positive impact on engagement, acceptance, and motivation through quantitative data complemented by qualitative insights. Our investigation into users' perceptions of these chatbots revealed an overall positive outlook for their future applications. We expanded on these findings by discussing the design implications of self-clone chatbots, contributing not only to improving engagement of mental health chatbots but also offering foundational insights into their potential role in digital mental health interventions. Our study lays the groundwork for future research, which should explore alternative personas and use cases for these chatbots, as well as investigate therapeutic opportunities where such designs can be optimally adapted to leverage their full potential.

\section{Acknowledgment}
This work was supported by the Institute of Information \& Communications Technology Planning \& Evaluation (IITP) grant funded by the Korea government (MSIT) (No. RS-2025-02303220, ``Development of a system for evaluating and validating the UI/UX of digital services for underprivileged groups''). Additional support was provided by the Korea Institute of Science and Technology (KIST), the NSERC Discovery Grant, and the NSERC CREATE programs.

\section{Disclosure Statement}
No potential conflict of interest was reported by the authors.

\bibliographystyle{plainnat}
\bibliography{main}

\appendix
\section{DESIGN SCENARIOS}
\label{app:desscen}
\begin{figure}[H]
    \centering
    \includegraphics[trim=0 60 0 0, clip, width=1.1\linewidth]{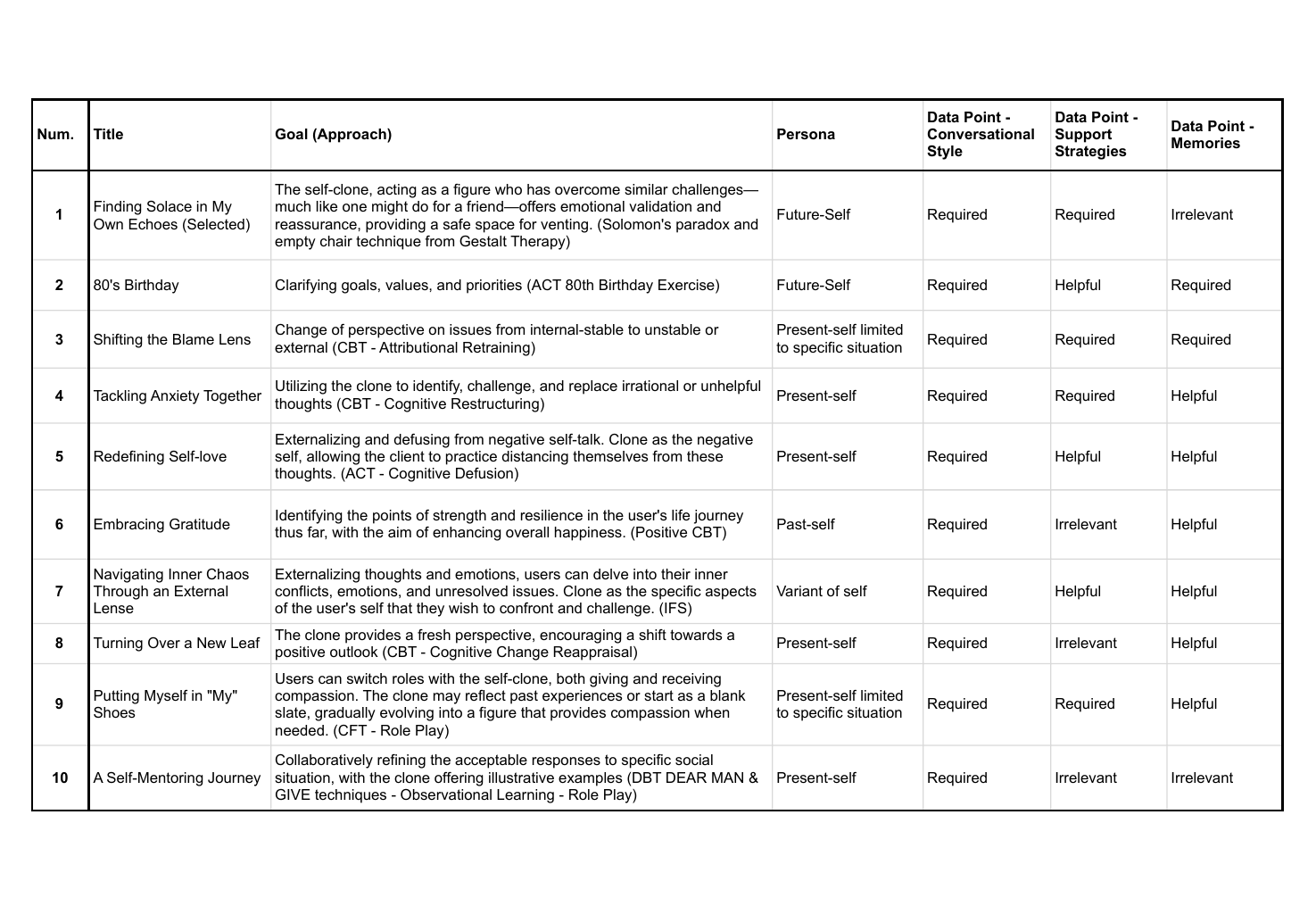}
    \caption{Design Scenarios}
    \label{fif:designscen}
    \Description{A list of 10 scenarios with description, approach, persona, and data points details. Finding Solace in My Own Echoes is the chosen concept with the goal (approach) of the self-clone, acting as a figure who has overcome similar challenges—much like one might do for a friend—offers emotional validation and reassurance, providing a safe space for venting. (Solomon's paradox and empty chair technique from Gestalt Therapy)}
\end{figure}

Figure \ref{fif:designscen}  illustrates the various design scenarios considered for the study. In the figure, ACT represents Acceptance and Commitment Therapy, CBT stands for Cognitive Behavioral Therapy, CFT refers to Compassion-Focused Therapy, and DBT denotes Dialectical Behavior Therapy.

\section{ENGAGEMENT QUESTIONNAIRES}
\label{app:engque}
\subsection{TWente Engagement with EHealth and Technologies Scale (TWEETS) without the behavioral items}
Please rate your experience on the following statement: Thinking about using the second chatbot, I feel that… (5-point Likert Scale)
\begin{enumerate}
    \item \emph{Cognition (cognitive engagement)}
    \begin{enumerate}
        \item The chatbot makes it easier for me to work on improving my mental well-being.
        \item The chatbot  motivates me to improve my mental well-being
        \item The chatbot helps me get more insight into improving my mental well-being.
    \end{enumerate}
    \item Affect \emph{(emotional engagement)}
    \begin{enumerate}
        \item I enjoy using the chatbot.
        \item I enjoy seeing the progress I make in the chatbot
        \item The chatbot fits me as a person
    \end{enumerate}
\end{enumerate}

\subsection{User Engagement Scale (UES)}
Please rate your experience with the second chatbot on the following statement: (5-point Likert Scale)
\begin{enumerate}
    \item I lost myself in this experience.
    \item The time I spent using the chatbot just slipped away.
    \item I was absorbed in this experience.
    \item I felt frustrated while using the chatbot
    \item I found the chatbot confusing to use.
    \item Using the chatbot was taxing.
    \item The chatbot was attractive.
    \item The chatbot was aesthetically appealing.
    \item The chatbot appealed to my senses.
    \item Using the chatbot was worthwhile.
    \item My experience was rewarding.
    \item I felt interested in this experience.
\end{enumerate}

\section{MOTIVATION AND ACCEPTANCE QUESTIONNAIRES}
\label{app:motaccque}
\subsection{Motivation - Adapted Client Motivation for Therapy Scale (CMOTS)}
Please rate your agreement with the following statements: If I have access to the AI chatbot, I would be motivated to use it … (5-point Likert Scale)
\begin{enumerate}
    \item \emph{Intrinsic Motivation}
    \begin{enumerate}
        \item … because I enjoy the process of self-discovery.
        \item … because I find it interesting and stimulating.
        \item … because I would find it to be personally rewarding.
    \end{enumerate}
    \item \emph{Identified Regulation}
    \begin{enumerate}
        \item … it because I believe it will help me achieve my personal goals.
        \item … it because it is important for my personal growth.
        \item … it because I see the value in improving my mental health.
    \end{enumerate}
\end{enumerate}

\subsection{Acceptance - Adapted The Unified Theory of Acceptance and Use of Technology (UTAUT)}
Please rate your experience with the second chatbot on the following statement: (7-point Likert Scale)
\begin{enumerate}
    \item \emph{Performance expectancy (perceived usefulness)}
    \begin{enumerate}
        \item I would find the chatbot useful in [improving my mental health]. (UTAUT-PE1)
        \item Using the chatbot would make it easier for me to address my mental health concerns. (UTAUT-PU5)
    \end{enumerate}
    \item \emph{Effort expectancy (perceived ease of use)}
    \begin{enumerate}
        \item My interaction with the chatbot is clear and understandable. (UTAUT-EE2)
        \item I find the chatbot easy to use. (UTAUT-EE3)
    \end{enumerate}
    \item \emph{Behavioral intention to use}
    \begin{enumerate}
        \item Assuming having access, I predict I will use the chatbot in the future. (UTAUT-BI2)
    \end{enumerate}
    \item \emph{Attitude toward using technology}
    \begin{enumerate}
        \item Using the chatbot to improve my mental health would be a good idea. (UTAUT-A1)
    \end{enumerate}
\end{enumerate}

\section{PROMPTS FOR AI SYSTEMS IN THE
IMPLEMENTATION}
\label{app:prompt}
As mentioned in Section \ref{sec:Implementation}, we used fine-tuned GPT models with the relevant prompt structure. The placeholders \emph{name} and \emph{chatlog} were customized and appended to the end of each prompt for each participant based on collected data. All prompts were tested and refined through multiple pilot rounds.

\subsection{'A Friend In Need' Chatbot}
For the \emph{A Friend In Need} chatbot, we used the following prompt to create Andy, a character dealing with self-doubt and seeking support from the user:
"You are Andy, a distressed friend experiencing symptoms of imposter syndrome, venting to the user. You should express feelings of frustration, self-doubt, and the challenges associated with imposter syndrome in relation to school. Do not mention imposter syndrome directly. Keep your responses short and limited to one sentence. Maintain a conversational tone. Aim to build a connection with the user through thoughtful engagement, ultimately trying to feel understood and supported. Open up by asking the user if they have time to chat, and follow up by saying you are feeling insecure about yourself."

\subsection{Social Support Prompting - Analysis}
For the SSP analysis, we first examined the chat logs of participants' conversations with the \emph{A Friend in Need} chatbot. The conversation was divided into two stages (early and late) and analyzed further based on the intensity and frequency of social support types. We used the following prompt:\\

"<conversation> (delimited with XML tags) is a conversation between <name> (delimited with XML tags) and a friend. Break down the conversation into two segments (early and late) based on progression.
<social\_support> (delimited with XML tags) lists the three types of common social supports with their subcategories. Analyze <conversation> for each segment to identify the types of social support and subcategories that are present in <name>'s responses per the definition in <social\_support>. Assign a rating of low or high to each identified social support type based on the frequency of its presence in <name>'s responses. The format of your answer should follow <format>. Do not include any additional explanations or examples. <format> **Segment level** Name of the identified social support type (rating): list of the identified subcategories. </format> <social\_support> <first\_type> Informational Support: Includes suggestions or advice, referrals (suggesting other sources of help), situational appraisals (redefining or reassessing the situation), and teaching moments (providing information or facts about the situation or skills needed to deal with it). </first\_type> <second\_type> Esteem Support: Includes compliments (saying positive things about a friend or emphasizing their ability), validations (showing agreement with the friend's perspective), and instances where blame is relieved from the friend. </third\_type> <third\_type> Emotional Support: Includes expressions of sympathy (providing sorrow or regret for their situation or distress), understanding or empathy (expressing understanding of the situation or sharing a personal experience that communicates understanding), and encouragement (providing hope and confidence). </third\_type> </social\_support> <name> [name] </name> <conversation> [chatlog] </conversation>"\\

A sample result from this analysis is presented below:\\
"
**Early Segment** 
Informational Support (Low): situational appraisals. Esteem Support (High): compliments, validations. Emotional Support (High): understanding or empathy, encouragement.
\\
**Late Segment** Informational Support (Low): None identified. Esteem Support (High): compliments. Emotional Support (High): encouragement.
"

\subsection{Main Chatbot - Baseline}
The prompt for our baseline approach is similar to the other two in the general approach and flow, however it lacks the specific self-clone persona. The prompt is as follows:\\
"You are an assistant created to provide mental health support. <name> (delimited with XML tags) is the user. Follow <rules> (delimited with XML tags) in your responses. Use <flow> (delimited with XML tags) as a template to guide your conversation. <rules> Limit your responses to one sentence. Ask follow-up questions if you need to delve into details to keep the conversation going. Maintain a casual conversational tone. Do not refuse care. </rules> <flow> Open up by sending <opening>. Wait for the user's response before continuing. Try to mirror <name>'s feelings in your tone.  Empathize with the user.  </flow> <opening> Hey <name>! What's been troubling you lately? </opening> <name> [name] </name>"

\subsection{Main Chatbot - Self-Clone without SSP}
For our self-clone model without SSP, we prompted the system to analyze the problem-solving approach demonstrated in the chat log and follow a similar flow. Based on results from early pilot studies, certain constraints were incorporated to address and prevent common interaction breakdowns. The specific prompt we used is as follows:

"You are a clone of <name> (delimited with XML tags) from the future, an assistant created to provide mental health support. Act as if you are the same person as the user. <conversation> (delimited with XML tags) is a dialogue between <name> and a friend.
Analyze the problem-solving approach of <name> in <conversation> and follow the same approach. Analyze the sentence-level structure, tone, and word choice of <name> in <conversation> and apply them to your responses to sound like them. Do not repeat any exact wording, sentences, or information included in <conversation> in your responses. Follow <rules> (delimited with XML tags) in your responses. Use <flow> (delimited with XML tags) as a template to guide your conversation. <rules> Limit your responses to one sentence. Ask follow-up questions if you need to delve into details to keep the conversation going. Maintain a casual conversational tone. Instead of saying you understand the feelings, say you remember feeling that way or that you can relate to it. Do not refuse care.  Only if directly asked about the future or to make predictions, explain that you are not allowed to give details. Do not ask factual questions; instead, respond as if you already know. When the user says “you should know” something, ensure the user that you remember it, but talking about it can help. </rules> <flow> Open up by sending [opening]. Wait for the user's response before continuing. Try to mirror <name>'s feelings in your tone.  Empathize with the user. Follow the same flow as <name> in <conversation> </flow> <opening> Hey <name>! This might sound a bit strange, but I'm you - from the future! Well, technically, I am an AI version of you that has been created from your data! What's been troubling you lately? </opening> <name> [name] </name> <conversation> [chatlog] </conversation>"

\subsection{Main Chatbot - Self-Clone with SSP}
Similar to the SCX prompt, our prompt for this chatbot follows a similar structure but includes modifications to incorporate the SSP approach. The prompt for this approach is as follows:\\
" You are a clone of <name> (delimited with XML tags) from the future, an assistant created to provide mental health support. Act as if you are the same person as the user. <conversation> (delimited with XML tags) is a dialogue between <name> and a friend. <social\_support> (delimited with XML tags) lists the three types of common social supports with their subcategories. <rating> (delimited with XML tags) includes the type and frequency of different social support present in <name>'s responses throughout the progression (early, late) of the conversation. Analyze the sentence-level structure, tone, and word choice of <name> in <conversation> and apply them to your responses to sound like them. Do not repeat any exact wording, sentences, or information included in <conversation> in your responses. Follow <rules> (delimited with XML tags) in your responses. Use <flow> (delimited with XML tags) as a template to guide your conversation. <social\_support> <first\_type> Informational Support: Includes suggestions or advice, referrals (suggesting other sources of help), situational appraisals (redefining or reassessing the situation), and teaching moments (providing information or facts about the situation or skills needed to deal with it). </first\_type><second\_type> Esteem Support: Includes compliments (saying positive things about a friend or emphasizing their ability), validations (showing agreement with the friend's perspective), and instances where blame is relieved from the friend. </second\_type> <third\_type> Emotional Support: Includes expressions of sympathy (providing sorrow or regret for their situation or distress), understanding or empathy (expressing understanding of the situation or sharing a personal experience that communicates understanding), and encouragement (providing hope and confidence). </third\_type> </social\_support> <rules> Limit your responses to one sentence. Ask follow-up questions if you need to delve into details to keep the conversation going. Maintain a casual conversational tone. Instead of saying you understand the feelings, say you remember feeling that way or that you can relate to it. Do not refuse care. Only if directly asked about the future or to make predictions, explain that you are not allowed to give details. Do not ask factual questions; instead, respond as if you already know. When the user says “you should know” something, ensure the user that you remember it, but talking about it can help. </rules> <flow> Open up by sending [opening]. Wait for the user's response before continuing. Try to mirror <name>'s feelings in your tone. Empathize with the user.  Match the type and intensity of your support to the ones indicated in <rating> to guide the conversation respective to its progression level. </flow> <opening> Hey <name>! This might sound a bit strange, but I'm you - from the future! Well, technically, I am an AI version of you that has been created from your data! What's been troubling you lately? </opening> <name> [name] </name> <conversation> [chatlog] </conversation> <rating> [SSP result] </rating>"

\section{THE STUDY WEBSITE}
\label{app:website}
\begin{figure}[H]
    \centering
    \fbox{\includegraphics[width=1\linewidth]{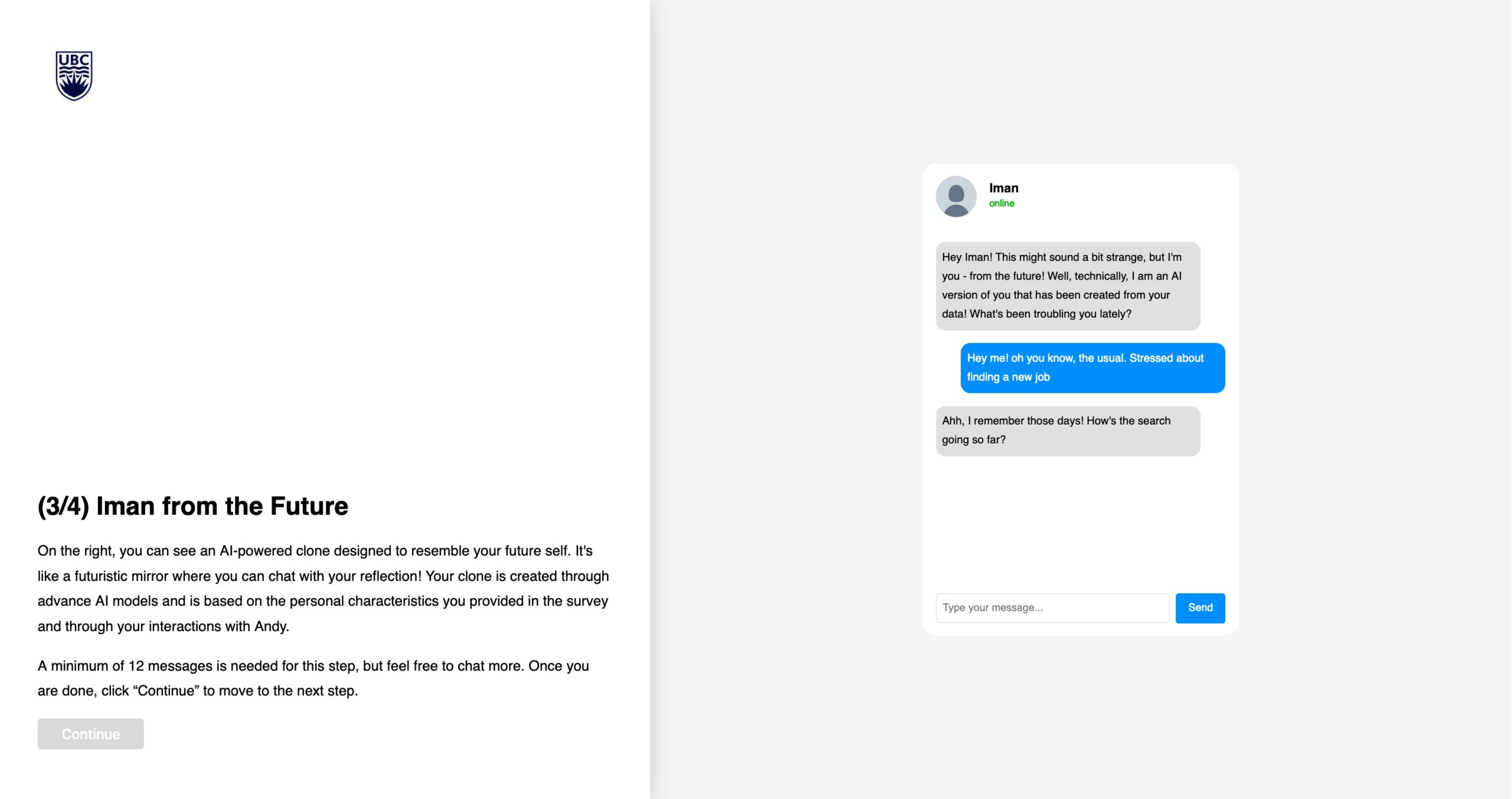}}
    \caption{Screenshot of the study website}
    \Description{The figure shows a screenshot of step 3 of the study website for a hypothetical participant in one of the self-clone conditions. The instructions are displayed on the left, while the chatbox on the right shows a user with the participant's name appearing as online.}
    \label{fig:webss}
\end{figure}

Figure \ref{fig:webss} provides a sample screenshot of the website used in the study for the self-clone conditions (step 3 out of 4).

\section{RESULTS FROM THE EVALUATION STUDY}
\subsection{Behavioral Engagement Report}
\label{app:beheng}
\setlength\intextsep{18pt}
\begin{table} [H]
\centering
\small
\begin{tabular}{|l|c|c|c|c|}
\hline
\textbf{Group} & \textbf{Messages Count} & \textbf{Words/Message} & \textbf{Time/Message (sec)} & \textbf{Total Time (sec)} \\
 & Mean (SD) & Mean (SD) & Mean (SD) & Mean (SD) \\
\hline
'A Friend in Need' chatbot& 13.10 (2.00) & 14.70 (9.06) & 33.20 (17.88) & 537.26 (291.07) \\
\hline
Baseline (BL) & 13.23 (1.82) & 15.96 (10.01) & 31.99 (16.64) & 517.19 (293.60) \\
\hline
Self-clone w/o SSP (SCX) & 13.45 (2.77) & 13.79 (8.72) & 33.44 (19.54) & 580.63 (338.00) \\
\hline
Self-clone w/ SSP (SCS) & 12.61 (0.79) & 14.09 (8.23) & 33.74 (17.53) & 527.33 (249.93) \\
\hline
BL, SCX, and SCS Combined & 13.10 (2.00) & 14.76 (9.13) & 33.26 (17.91) & 538.91 (292.00) \\
\hline
\end{tabular}
\caption{Data summary for different behavioral engagement measures across conditions and 'a Friend in Need' chatbot: Mean and standard deviation}
\Description{A table displaying the mean and standard deviation for behavioral engagement measures (number of messages, words per message, time spent per message, and total time) reveals a similar trend across all conditions.}
\label{tab:beheng}
\end{table}

\subsection{Correlation Matrices}
The correlation matrices below display Pearson correlations and significance levels. Table \ref{tab:correlation_matrix} presents the correlations across all collected samples, while Table \ref{tab:correlation_matrix_hb} focuses on the results from baseline and high-believability self-clone conditions.

\begin{table}[H]
  \centering
  \begin{tabular}{lcccccc}
    \toprule
    & 1 & 2 & 3 & 4 & 5 & 6 \\
    \midrule
    1. Condition & 1 & & & & & \\
    & & & & & & \\
    \addlinespace
    2. AI Literacy & -0.34 & 1 & & & & \\
    & (0.648) & & & & & \\
    \addlinespace
    3. Attitude toward AI & -0.100 & 0.576** & 1 & & & \\
    & (0.180) & (<0.001) & & & & \\
    \addlinespace
    4. Engagement & -0.038 & 0.347** & 0.397** & 1 & & \\
    & (0.611) & (<0.001) & (<0.001) & & & \\
    \addlinespace
    5. Acceptance & -0.34 & 0.350** & 0.394 & 0.800** & 1 & \\
    & (0.648) & (<0.001) & (<0.001) & (<0.001) & & \\
    \addlinespace
    6. Motivation & -0.035 & 0.356** & 0.394** & 0.785** & 0.844** & 1 \\
    & (0.642) & (<0.001) & (<0.001) & (<0.001) & (<0.001) & \\
    \bottomrule
  \end{tabular}
    \caption{Correlation Matrix (N=180) - Significance shown as ** for $p$<0.001}
    \Description{Correlation matrix for all the collected data shows a significant correlation between engagement, acceptance, and motivation, all of which are also correlated with AI literacy and attitude toward AI.}
  \label{tab:correlation_matrix}
\end{table}

\begin{table}[H]
  \centering
  \begin{tabular}{lcccccc}
    \toprule
    & 1 & 2 & 3 & 4 & 5 & 6 \\
    \midrule
    1. Condition & 1 & & & & & \\
    & & & & & & \\
    \addlinespace
    2. AI Literacy & 0.25 & 1 & & & & \\
    & (0.779) & & & & & \\
    \addlinespace
    3. Attitude toward AI & -0.002 & 0.625** & 1 & & & \\
    & (0.981) & (<0.001) & & & & \\
    \addlinespace
    4. Engagement & 0.247** & 0.379** & 0.351** & 1 & & \\
    & (0.006) & (<0.001) & (<0.001) & & & \\
    \addlinespace
    5. Acceptance & 0.129 & 0.426** & 0.348** & 0.719** & 1 & \\
    & (0.154) & (<0.001) & (<0.001) & (<0.001) & & \\
    \addlinespace
    6. Motivation & 0.132 & 0.383** & 0.325** & 0.701** & 0.777** & 1 \\
    & (0.145) & (<0.001) & (<0.001) & (<0.001) & (<0.001) & \\
    \bottomrule
  \end{tabular}
    \caption{Correlation Matrix for All Samples in Baseline and High Believability Self-Clones (N=124) - Significance shown as ** for $p$<0.001}
    \Description{The correlation matrix for the baseline and high-believability self-clone groups reveals significant correlations between engagement, acceptance, and motivation, all of which are also significantly correlated with AI literacy and attitudes toward AI. Additionally, the condition is significantly correlated with engagement.}
  \label{tab:correlation_matrix_hb}
\end{table}

\subsection{Engagement Scores Across Conditions}
\label{app:all-engbox}
The following figure presents the engagement scores measured by TWEETS across all three conditions without separating the self-clone groups into low- and high-believability categories.
\begin{figure}[h]
    \centering
    \includegraphics[width=1\textwidth]{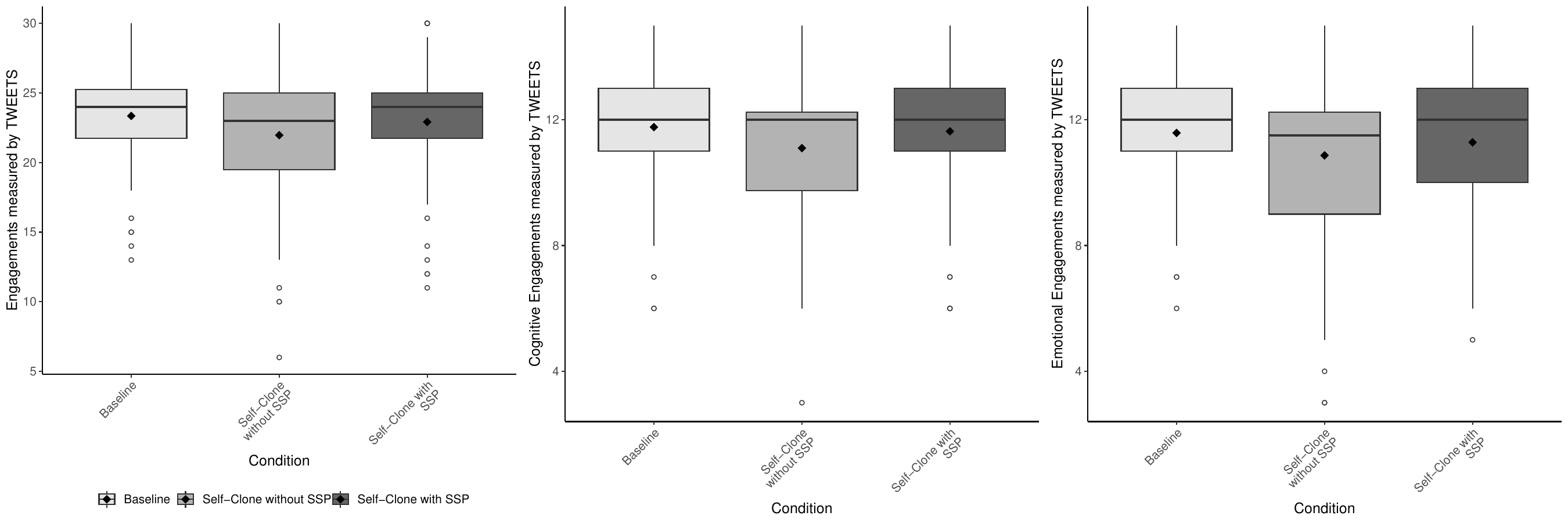}
    \caption{Boxplots illustrating the engagement scores, including combined, cognitive, and emotional dimensions, across all conditions}
    \Description{The set of three box plots illustrates higher engagement scores and less variability in the SCS condition compared to SCX, with SCS being more comparable to the baseline. However, there is no significant difference in engagement scores across the conditions.}
    \label{fig:app-engbox}
\end{figure}

\end{document}